\documentclass[journal]{IEEEtran}
\usepackage{graphicx,cite,soul,framed,epsfig,mathptmx,
amsmath,amssymb,color,flushend,gensymb,fancyhdr,subfigure}
\usepackage{lipsum}
\definecolor{shadecolor}{rgb}{1,0.8,0.3}

\ifCLASSINFOpdf
\else
\fi
\newtheorem{remark}{Remark}

\begin{document}
\title{Multi-Objective $H_{\infty}$ Control for String Stability of Cooperative Adaptive Cruise Control Systems}
\author{Erkan~Kayacan~\IEEEmembership{} 
\thanks{The author is with Coordinated Science Laboratory, University of Illinois at Urbana-Champaign, Urbana, Illinois 61801, USA. e-mail: {\tt\small erkank@illinois.edu }} }

\markboth{\textbf{PREPRINT VERSION:} IEEE TRANSACTIONS ON INTELLIGENT VEHICLES,  Volume 2, Issue 1, 2017.}{Shell \MakeLowercase{\textit{et al.}}: Bare Demo of IEEEtran.cls for Journals}

\maketitle
\begin{abstract}
Autonomous vehicle following systems are playing a decisive role to increase vehicle density on roads by shortening inter-vehicle time gaps. However, disturbance attenuation along a platoon of vehicles, i.e., string stability, is being a challenging task while time gap is getting shorter. In order to guarantee the string stability of a vehicle platoon, a multi-objective $H_{\infty}$ control formulation for adaptive cruise control and cooperative adaptive cruise control structures has been investigated in this paper. The proposed control method solves an optimization problem and achieves a controller that is able to provide not only the system stability, but also the string stability as distinct from the traditional $H_{\infty}$ control. 
\end{abstract}

\IEEEpeerreviewmaketitle

\section{Introduction}

\begin{figure*}[t!]
\begin{center}
  \includegraphics[width=6.8 in] {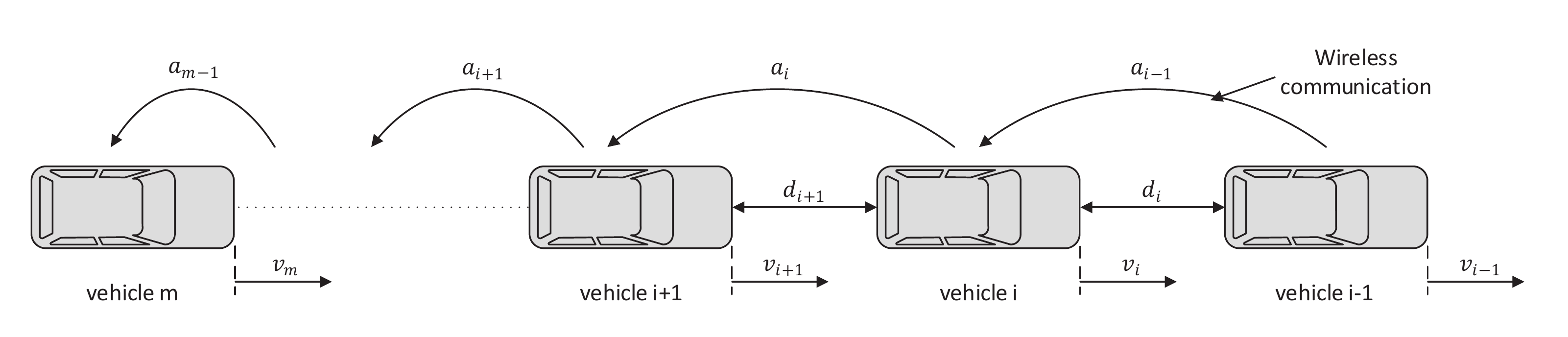}\\
  \caption{Schematic of a platoon of vehicles}\label{schematic}
   \end{center}
\end{figure*}

\IEEEPARstart{A}{daptive} cruise control (ACC) systems, extension of cruise control systems, are so widespread and commercially available in vehicle market at the present time. A survey indicated that $14 \%$ of U.S. drivers had ACC system in their vehicles and $30 \%$ of newly sold cars were equipped with ACC in 2014 \cite{Ashley2016}. They regulate the vehicle speed if there is a preceding vehicle driven at a lower speed or inter-vehicle distance is smaller than predefined spacing policy. In such systems, a radar is used to measure inter-vehicle distance and its rate of change to detect any preceding vehicle. In former commercial applications, ACC sytems were considered for very large inter-vehicle distances as a comfort system rather than a safety system \cite{Vahidi2003}. 

In the last decades, a vast majority of people prefer to use their private cars increasingly resulting in a fact that our mobility is jeopardized by traffic congestion. In order to be able to use the existing road capacity in a more efficient way, a possible solution is to increase vehicle density by decreasing inter-vehicle distances considering safety issues. Since shorter inter-vehicle distances may cause dangerous situations for human drivers, longitudinal speed control is a requirement. In this instance, ACC systems have been recognized to be used in future commercial applications. Nowadays, researchers have focused on inter-vehicle data exchange through wireless communications to further extend the capabilites of the traditional ACC systems. The extended version of ACC is known as cooperative ACC (CACC) and ensures smaller time gaps than the traditional ones \cite{Rajamani2002}.  

As ACC and CACC systems are employed to ensure short inter-vehicle distances, disturbance amplifications in upstream direction may occur. For this reason, disturbance attenuation, also known as the string stability, is an essential issue for vehicle platoons as well as the system stability \cite{Ploeg2014Lp}. A vehicle platoon is string stable if changes in the speed of vehicles are decreasing along the vehicle string. 

Numerous types of controllers and control structures have been proposed for ACC and CACC systems in literature. Conventional, e.g., propotional-derivative, controllers have been proposed in  \cite{Naus2010,Ploeg2011,Guvenc2012}. However, they cannot take the string stability requirement into account during the controller design stage. Therefore, well-tuning of the controller is neccessary to achieve string stable behaviour. In order to decrease inter-vehicle distances, leader$\&$preceding communication structure has been proposed in \cite{Milanes2014}. In this structure, the error and data exchange between not only the preceding and following vehicles but also the leader and following vehicles exist. Thus, there are two feedback controllers: preceding gap error and leading gap error controllers. 

A model predictive control (MPC) algorithm, which can deal with constraints on states and inputs, has been formulated as a feedback controller in \cite{Fanping2010} for ACC systems. A quadratic cost function consisting of the gap and speed errors between the preceding and following vehicles is minimized. In real-time, it has been observed that the controller has given satisfactory performance, while the speed of the preceding car has been constant. However, the system has not been able to reach the desired time gap, while the preceding car has been accelerated or decelerated. A similar MPC consisting of the gap and speed errors has been formulated and a comparable performance has been observed in real-time \cite{Kianfar2012}. Moreover, a reinforcement learning control method has been applied as a model-free control method in \cite{Desjardins2011}. The advantage of this method is to be able to control the system without knowing its mathematical model, because the modeling and identification of these systems are time-consuming. In addition, these processes are always subjected to uncertainties. As observed from some works in literature, there exists no unique control design method that guarantees both the system stability and string stability. Moreover, these controllers are decentralized ones, which calculate their own feedback control actions without knowing their predecessors' position and speed errors. Another well known control structure is centralized one that uses the information of all the vehicles in the string \cite{Liang1999}. 

In addition to decentralized control structures of vehicle platoons, distributed ones have been suggested to increase control performance. They require less computation time when compared to their centralized counterparts \cite{erkanDiNMPC, erkanDeNMPC}. Distributed model predictive control for vehicles with nonlinear dynamics has been formulated in \cite{Dunbar2012} and a distributed particle swarm optimization method has been proposed in \cite{Liu2015}. In these control structures, the controllers calculate their own inputs and receive the calculated position and speed error trajectories from their predecessors in addition to their inputs. In such structures, the required transmission results in overwhelming communication data exchange. Moreover, the communication delay has not been taken into consideration. 

This paper focuses on a decentralized control approach due to its relevance to real world implementations in everyday traffic in which the string stability requirement is imposed as a controller design requirement. The main contribution of this paper is to formulate a multi-objective $H_{\infty}$ control to guarantee the string stability of a vehicle platoon in ACC and CACC cases due to the fact that it is naturally fit to the $H_{\infty}$ norm string stability condition for linear systems as distinct from the previous studies \cite{Hao2013,Ploeg2014Synthesis}. Moreover, $H_{\infty}$ control allows to make trade-offs between vehicle following performance, system robustness and string stability.  

This paper is organized as follows: The spacing policy and the string stability requirements are defined in Section \ref{ProblemFormulation}. Control structures are presented in Section \ref{ControlStructure}. Traditional and multi-objective $H_{\infty}$ controllers are designed in Section \ref{controllerdesign}. The simulation results are given in Section \ref{SimulationStudies}. Finally, some conclusions are drawn from this study in Section \ref{Conclusions}.

\section{Problem Formulation}\label{ProblemFormulation}

The main aim in (C)ACC systems is to keep vehicles as close as possible to increase the vehicle density on highways and to prevent the amplification of disturbances throughout the string therewithal. The latter is known as the string stability of vehicle platoons \cite{oncu2014,pirani2016graph}. When conventional controllers are designed for such systems, decreasing inter-vehicle relative distance can result in string instability. Moreover, (C)ACC systems maintain accurate and smooth speed of the vehicles despite disturbances.

\subsection{Spacing Policy}

The vehicles in a platoon are connected with each other through the spacing policy. In other words, each vehicle has to follow its preceding vehicle with a desired relative distance defined by the spacing policy. In general, there are three types of spacing policies: constant distance, constant time gap and variable time gap policies \cite{swaroop1994,Wang2004}. The constant time gap policy was used in former ACC implementations when the vehicle density on highways was not critical and the variable time gap policy was proposed to guarantee traffic-flow stability. The former policy leads to linear control law while the latter policy leads to nonlinear control law. Nowadays, the constant time gap policy has been used to lead to increased traffic capacity by taking vehicle dynamics into account \cite{Jing2005}. In this study, our aim is to lead to higher traffic-flow utilization of the highway; therefore, the constant time gap policy, i.e., a velocity dependent spacing policy, is considered.

In Fig. \ref{schematic}, it is denoted that $d_{r,i}$ is the desired distance between the vehicle $i$ and its preceding vehicle $i-1$, $d_{0,i}$ is the standstill distance, which is the required distance between stopped vehicles, $l_{i}$ is the length of the vehicle $i$, $a_{i}$ is the acceleration of the vehicle $i$ and $m$ is number of following vehicles in a platoon. The time-headway spacing policy to follow the preceding vehicle with a desired relative distance is generally defined as \cite{Wang2004}:
\begin{eqnarray}\label{spacingpolicy}
d_{r,i}= d_{0,i} + h_{i} v_{i}, \;\;\;\;\; 1 \leq i \leq m
\end{eqnarray}
where $d_{r,i}$ is the desired relative distance between the front bumper of the $i$th vehicle to its predecessor's rear bumper, $d_{0,i}$ is the standstill distance, $h_{i}$ is the desired time headway, representing the time that it will take the $i$th vehicle to arrive at the same position as its predecessor when the standstill distance $d_{0,i}=0$ is equal to zero. As can be seen in \eqref{spacingpolicy}, the spacing policy consists of a constant and a velocity dependent parts.

In this paper, the desired standstill distance $d_{0, i}$ is regarded as an extension of the vehicle length $L_{i}= l_{i} + d_{0,i}$. Thus, the desired standstill distance is neglected in the formulation so that the desired relative distance is written as follows:
\begin{eqnarray}\label{spacingpolicy2}
d_{r,i}= h_{i} v_{i}, \;\;\;\;\; 1 \leq i \leq m
\end{eqnarray}
For $h_{i} > 0$, the spacing policy is called the constant time-headway time policy which aims a constant inter-vehicle time gap.

The actual relative distance is written as follows:
\begin{eqnarray}\label{relativedistance}
d_{i}=p_{r,i-1} - p_{f,i} 
\end{eqnarray}
where $p_{r, i-1}$ and $p_{f,i}$ are respectively the positions for the rear bumper of the vehicle $i-1$ and the front bumper of the vehicle $i$.

The relative spacing error between the front bumper of the $i$th vehicle to its predecessor's rear bumper is defined by taking the spacing policy in \eqref{spacingpolicy2} into account as follows:
\begin{eqnarray}\label{relativeerror}
e_{i}=p_{r,i-1} - p_{f,i}  - h_{i} v_{i}
\end{eqnarray}

\subsection{String Stability}

The stability of dynamical systems is evaluated with the system states over time, whereas the string stability focuses on the propagation of the system responses along a cascade of systems.

The traditional methods are based on Lyapunov stability \cite{Rajamani2002}. In these analysis, the string stability is proved based-on asymptotic stability of interconnected vehicles by focusing on perturbations on initial conditions. In \cite{sheikholeslam1992}, the response to the perturbation on the initial condition of a vehicle is taken into account. However, the perturbations on the initial conditions of the other vehicles in a platoon and external disturbances to a platoon are neglected. Therefore, these approaches are not sufficient to guarantee the string stability due to the fact that changes in the velocity of the leading vehicle can occur.

The infinite length string of interconnected vehicles is also investigated to prove the string stability of a platoon \cite{Santhanakrishnan2003}. In this approach, it is assumed that the vehicles are identical and the model is formulated in a state space form. After taking the bilateral Z-transform, it is executed over the vehicle index. String stability can be determined by looking at the eigenvalues of the state matrix. However, the string stability of infinite-length vehicles may not be able to converge to the string stability of finite-length vehicles as there will be always first and last vehicles, whose dynamics may be different from vehicle dynamics in the platoon. Therefore, this approach is not an appropriate candidate to prove the string stability.

A performance-based approach for string stability of a platoon is proposed with and without leading vehicle information for linear cascaded system \cite{swaroop1994}. This approach is extended for exchanging information between vehicles of a platoon \cite{swaroop1999}. The effects of communication delay are investigated in \cite{Liu2001}. A decentralized controller is designed by decoupling the interactions between vehicles and the limitations on performance are analyzed \cite{Seiler2004}. In this approach, the string stability is determined by considering amplification of signals such as the relative distance error, the velocity, the acceleration or the control input in the platoon while the vehicle index increases.

In this paper, the method in \cite{Seiler2004} is used to design a controller. The string stability is quantified by the magnitude of the string stability transfer function as
\begin{eqnarray}\label{stringstability}
SS_{\Delta_{i}} (s)=\frac{\Delta_{i} (s)}{\Delta_{i-1} (s)}, \;\;\;\;\; 1 \leq i \leq m
\end{eqnarray}
where $\Delta$ is the interested signal.

A condition on the maximal amplification of perturbation along string in a platoon can be considered as the requirement of the string stability \cite{Ploeg2014Lp}. The maximal amplification can be written by the $H_{\infty}$ norm of the string stability transfer function in \eqref{stringstability} as follows:]
\begin{eqnarray}\label{stringstability2}
\| SS_{\Delta_{i}} (j \omega) \|_{\infty}= \sup_{\omega \in \mathcal{R}}  \| SS_{\Delta_{i}} (j \omega) \| , \;\;\;\;\; 1 \leq i \leq m
\end{eqnarray}
where $\mathcal{R}$ is the set of real numbers. The following condition can be formulated by considering \cite{SSCACC}
\begin{eqnarray}\label{stringstability3}
\sup_{\omega \in \mathcal{R}}  \| SS_{\Delta_{i}} (j \omega)  \|  \leq 1 \;\;\; \forall \omega, \;\;\;\;\; 1 \leq i \leq m
\end{eqnarray}

\subsection{Longitudinal Vehicle Dynamics}
In longitudinal vehicle dynamics, it is assumed that the input of the vehicle is the throttle system while the output of the vehicle is the position. Therefore, a model is required to describe the relation between the desired acceleration to the throttle system and the position of the vehicle. The third-order linear model is proposed to represent the longitudinal dynamics for each vehicle in a platoon as follows \cite{Xiao2011}:
\begin{eqnarray}\label{longitudinal}
\dot{p}_{i} (t) & = & v_{i} (t) \nonumber \\
\dot{v}_{i} (t) & = & a_{i} (t) \nonumber \\
\dot{a}_{i} (t) & = & -\tau_{i}^{-1} a_{i} (t) + \tau_{i}^{-1} u_{i} (t-\phi_{i})
\end{eqnarray}
where $p_{i}$, $v_{i}$, $a_{i}$ are respectively the position, the velocity and the acceleration, $\tau_{i}$ is time-lag, $u_{i}$ is the input of the system, i.e., the desired acceleration, $\phi_{i}$ is the actuator time-delay between the desired acceleration and the applied acceleration for the vehicle $i$. This linear model is a simpler version of the detailed nonlinear model due to non-linear dynamics of engine and drive train, aero- dynamic drag, and rolling resistance, and obtained by using input-output linearization method \cite{Sheikholeslam1993,Stankovic2000}.

The longitudinal vehicle dynamics of the vehicle $i$ in \eqref{longitudinal} can be written in a transfer function form by using the Laplace transform as follows:
\begin{eqnarray}\label{longitudinaltf}
G_{i} (s) = \frac{1}{s^2 (\tau_{i} s +1)}e^{-\phi_{i} s}, \;\;\;\;\; 1 \leq i \leq m
\end{eqnarray}

It is to be noted that all vehicles in a platoon are assumed to be identical in this study.

\section{Existing Control Structures} \label{ControlStructure}

\subsection{ACC Control Structure}

ACC employs forward-looking sensors, i.e., radar, for longitudinal dynamics control of the vehicle by using throttle and brake actuations \cite{Caveney2010}. The main objective of its structure is to follow the preceding vehicle at a desired relative distance $d_{r,i}$. The relative distance $d_{i}=p_{r,i-1} - p_{f,i}  $ and the relative velocity $\dot{d}_{i}=v_{i-1}-v_{i}$ are measured by using a radar. In traditional ACC systems, a feedback controller is used to control the spacing error $e_{i}=d_{i}-d_{r,i}$ between the actual and desired distances. A positive control action, i.e., acceleration, is required when the spacing error is positive. Moreover, a negative control action, i.e., braking, is required when the spacing error is negative.

\begin{figure}[t!]
\begin{center}
  \includegraphics[width=3 in]{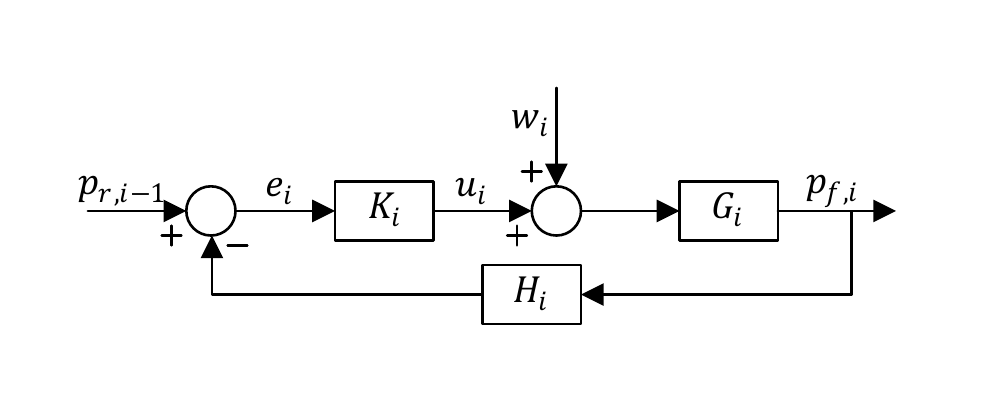}\\
  \caption{ACC control structure}\label{control_structure_ACC}
   \end{center}
\end{figure}

The transfer function of the velocity-dependent spacing policy is written by taking \eqref{relativeerror} into account as follows:
\begin{eqnarray}\label{H_tf}
H_{i} (s) = h_{i} s +1 , \;\;\;\;\; 1 \leq i \leq m
\end{eqnarray}
Considering the velocity dependent spacing policy, the ACC control structure is illustrated in Fig. \ref{control_structure_ACC}. The disturbance is denoted by $w_{i}(t)$ for the vehicle $i$.

The closed-loop sensitivity transfer function and the closed-loop complementary sensitivity transfer function of ACC systems are written below:
\begin{eqnarray}\label{eq_ACC_S}
S_{i} (s) & = & \frac{1}{1+ G_{i} (s) H_{i} (s) K_{i} (s)} , \;\;\;\;\; 1 \leq i \leq m \\ \label{eq_ACC_T}
T_{i} (s) & = & \frac{G_{i} (s) K_{i} (s)}{1+ G_{i} (s) H_{i} (s) K_{i} (s)} , \;\;\;\;\; 1 \leq i \leq m
\end{eqnarray}

\subsection{CACC Control Structure}

CACC systems are the extended versions of ACC systems and a reliable communication is crucial so that it is realized without any central management in networks consisting of vehicles in a platoon \cite{Wischhof2005}. The reason is that all vehicles in a platoon must be within each other's communication range in a centralized case. This means that a platoon with 20-25 vehicles can be a realistic assumption when the IEEE 802.11p communication protocol is used \cite{Vinel2015}. The communication bandwidth becomes insufficient when the number of vehicles increases in a platoon. For this reason, short range wireless communications are a better option for autonomous vehicle technology so that dedicated short range communication has been chosen as the standard communication protocol for cooperative vehicle safety applications \cite{Dey2015}. Moreover, there is another potential limitation for platoon length in entering and exiting ramps on highways. Consequently, a decentralized controller design is focused in this paper by taking the feasibility of implementation into account.

\begin{figure}[t!]
\begin{center}
  \includegraphics[width=3.4 in]{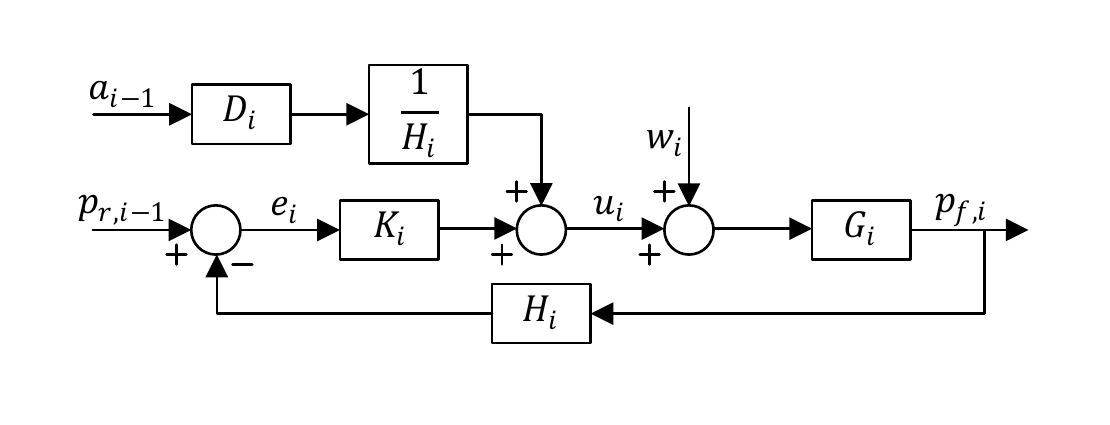}\\
  \caption{CACC control structure}\label{control_structure_CACC}
   \end{center}
\end{figure}

In CACC systems, the obtained acceleration of the preceding vehicle $a_{i-1}$ by wireless communication is used as a feedforwad control action for the following vehicle $i$. If there is no communication, the system becomes a traditional ACC system. The CACC control structure is illustrated in Fig. \ref{control_structure_CACC}. As can be seen from this figure,  since the acceleration is obtained through wireless communication, it includes a communication delay $D_{i} (s)$. A feedforward filter $1 / H_{i}(s)$ is also used. The communication delay transfer function is written as
\begin{equation}\label{Dtimedelay}
D_{i} (s) = e^{-\theta_{i} s}, \;\;\;\;\; 1 \leq i \leq m
\end{equation}
where $\theta_{i}$ is the constant time-delay.

The closed-loop sensitivity transfer function and the closed-loop complementary sensitivity transfer function are written as follows:
\begin{eqnarray}\label{eq_CACC_S}
S_{i} (s) &=& \frac{1 - D_{i} (s) }{1+ G_{i} (s) H_{i} (s) K_{i} (s)} , \;\;\;\;\; 1 \leq i \leq m \\ \label{eq_CACC_T}
T_{i} (s) &=& \frac{D_{i} (s) + G_{i} (s) K_{i} (s) H_{i} (s)}{H_{i} (s) (1+ G_{i} (s) H_{i} (s) K_{i} (s))}, \;\;\;\;\; 1 \leq i \leq m
\end{eqnarray}

\section{$H_{\infty}$ Control Structures}\label{controllerdesign}
\subsection{Traditional $H_{\infty}$ Control  }

The aim of the traditional $H_{\infty}$ control method is to calculate a controller $K$ that minimizes the $H_{\infty}$ norm of the closed loop system.  The traditional optimal design problem is formulated as follows:
\begin{eqnarray}\label{tra_Hinf}
\displaystyle\min\limits_{K \in \mathcal{K}} && \Gamma \nonumber \\
\textrm{subject to} && \| W_{S} S \|_{\infty} \leq \Gamma \nonumber \\
                    && \| W_{U} U \|_{\infty} \leq \Gamma \nonumber \\
                    && \| W_{T} T \|_{\infty} \leq \Gamma 
      \end{eqnarray}
where $\mathcal{K}$ is the set of stabilizing controllers, $K$ is the parametric model of the controller, $W_{S}$, $W_{U}$ and $W_{T}$ are respectively weighting functions which penalize respectively the error signal, control signal and output signal, $\Gamma$ is the best achieved value for the closed-loop $H_{\infty}$ norm and must be smaller than $1$. A Riccati-based approach and a linear-matrix-inequality-based approach were developed to solve this optimization problem \cite{Zhou198885,Gahinet1994}. $H_{\infty}$ norm has limited practical value and requires additional weighting functions. In general, since determining proper weighting functions is an iterative procedure, it is a tiresome task. The reason is that all control objectives and constraints are merged into one objective in \eqref{tra_Hinf}. For this reason, it is well known that $H_{\infty}$ control synthesis is not an easy and intuitive control design method. 

The general $H_{\infty}$ control method is illustrated in Fig. \ref{P_tra_fig}. The augmented plant $P$ is the system to be controlled, and has two inputs and two outputs: $r$ is the exogenous input, $u$ is the control signal, $y_{1}$ is the exogenous output and $y_{2}$ is the controller input. In other words, the error signal $y_{1}$ is minimized, and the measured variable $y_{2}$ is used in $K$ to calculate the manipulated variable $u$. All system is formulated as follows:
\begin{eqnarray}\label{Hinf_eq}
\left(
  \begin{array}{c}
  y_{1} \\
  y_{2} \\
  \end{array}
  \right) = P
  \left(
  \begin{array}{c}
  r  \\
  u \\
    \end{array}
  \right) 
\end{eqnarray}
where 
\begin{eqnarray}
P = \left(
  \begin{array}{c|c}
  W_{S} & -W_{S} G \\
    0   & -W_{U} \\
    0   & W_{T} G \\
    \hline
    I   & -G \\
  \end{array}
  \right)
\end{eqnarray}

The controller $u$ is calculated as 
\begin{equation}\label{Hinf_u_eq}
u=K  y_{2}
\end{equation}
where $K$ is parameterized as a dynamical system and has the same order of the augmented plant $P$.

\subsection{Multi-Objective $H_{\infty}$ Control }

Dividing the classical $H_{\infty}$ control problem into a multi-objective $H_{\infty}$ control problem makes the weighting function design easier and possible to add different objectives. As can be observed from previous subsection, the traditional $H_{\infty}$ control structure has only one objective and it has not been able to include the string stability of the CACC system regarding the chosen spacing policy. In this subsection, an additional constraint to satisfy the string stability condition in \eqref{stringstability3} is added on the complementary sensitivity function to guarantee the string stability. The string stability condition is written as follows:
\begin{equation}\label{}
  \| T \|_{\infty} \leq  1
\end{equation}
where $T$ is the complementary function and formulated respectively in \eqref{eq_ACC_T} and \eqref{eq_CACC_T} for ACC and CACC systems.

The multi-objective $H_{\infty}$ control problem for (C)ACC systems is written as follows:
\begin{eqnarray}\label{revied_Hinf_ACC}
\displaystyle\min\limits_{K \in \mathcal{K}, \Gamma_{S}, \Gamma_{U}, \Gamma_{T} } && \Gamma_{S} + \Gamma_{U} + \Gamma_{T} \nonumber \\
\textrm{subject to} && \| W_{S} S \|_{\infty} \leq \Gamma_{S} \nonumber \\
                    && \| W_{U} U \|_{\infty} \leq \Gamma_{U} \nonumber \\
                    && \| W_{T} T \|_{\infty} \leq \Gamma_{T} \nonumber \\
                    && \| T \|_{\infty} \leq 1
\end{eqnarray}
If $\Gamma_{S}$, $\Gamma_{U}$ and $\Gamma_{T}$ $<$ $1$, the sufficient conditions are satisfied for the controller design. The optimization problem above is a multi-objective control problem consisting of multiple $H_{\infty}$-norms on closed-loop subsystems. This allows seperated design objectives and constraints for each  selected closed-loop subsystem. However, the control problem requires bilinear matrix equalities that are non-convex and hard to solve.  As solution strategies, the Lyapunov shaping and G-shaping paradigms have been proposed in \cite{scherer1997,Oliveira2002}.  Moreover, one of the major issues with frequency domain analysis is that it cannot deal with constraints on state and control variables unlike model predictive control.

\begin{figure}[h!]
\centering
\subfigure[ ]{
\includegraphics[width=2.15in]{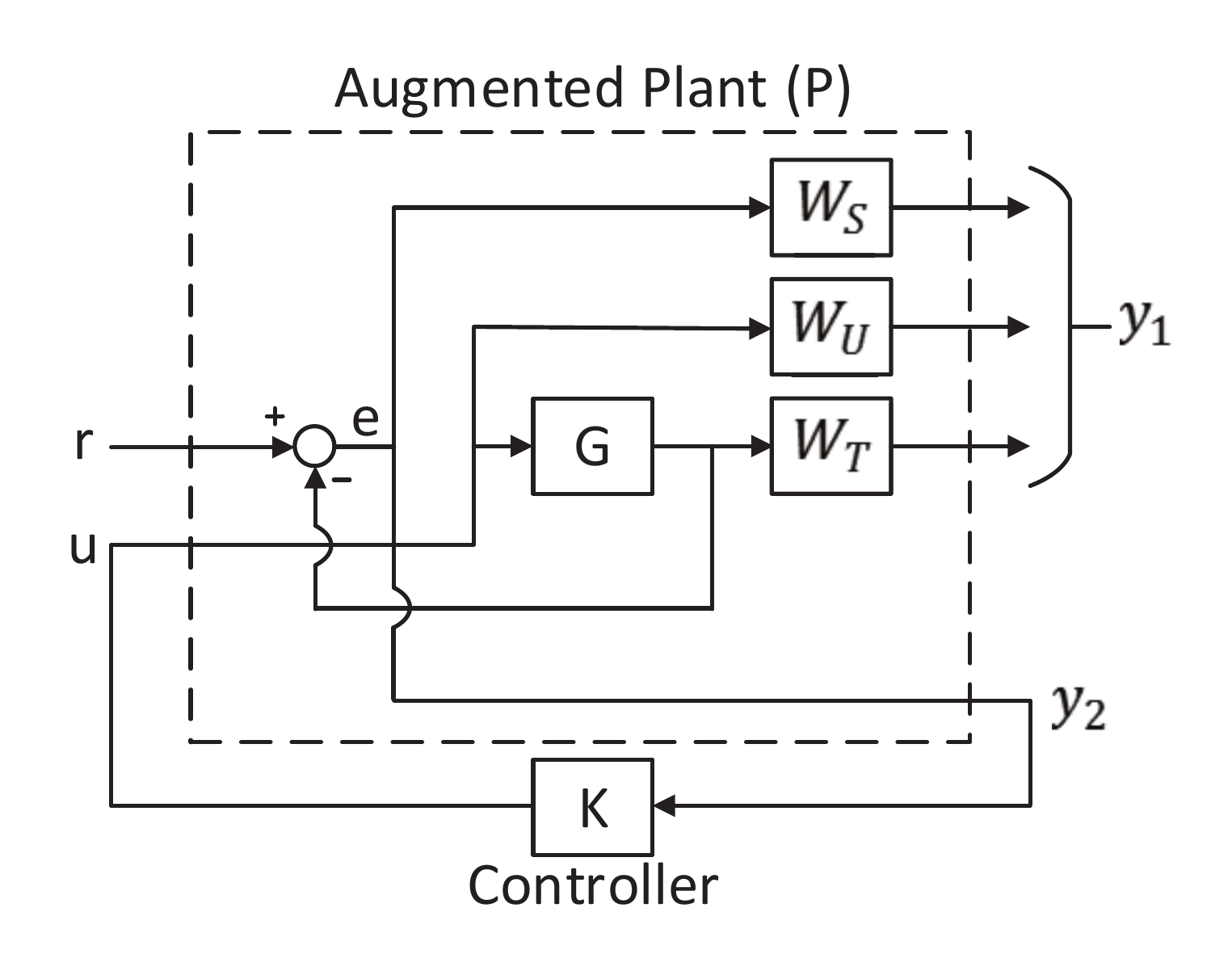}
\label{P_tra_fig}
}
\subfigure[ ]{
\includegraphics[width=2.15in]{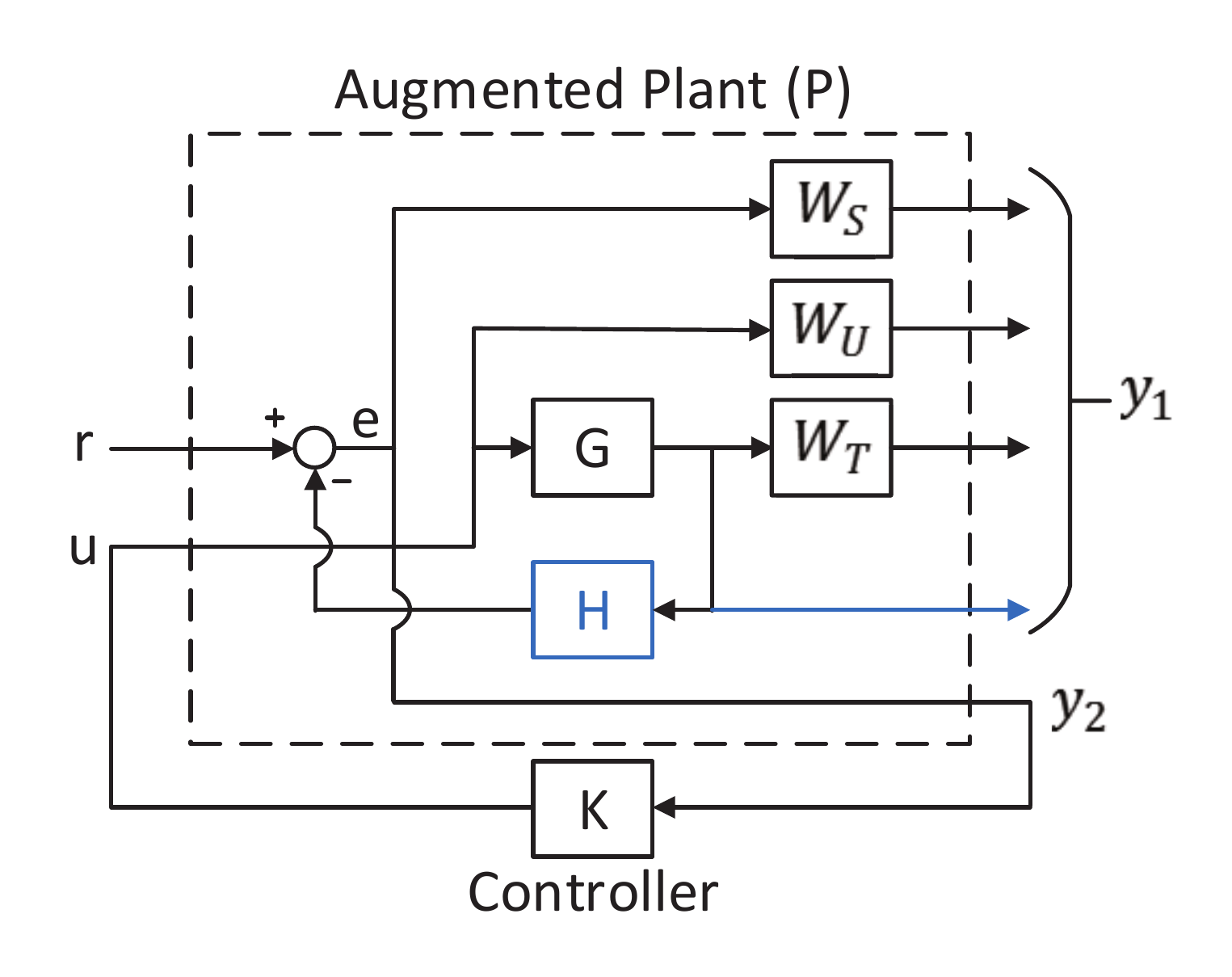}
\label{P_new2_fig}
}
\caption[Optional caption for list of figures]{(a) Traditional augmented plant (b) Augmented plant for ACC systems in multi-objective $H_{\infty}$ control }
\label{sensors}
\end{figure}

In addition to adding string stability condition on the problem formulation, the traditional $H_{\infty}$ control structure does not contain the spacing policy. For this reason, the control structure is re-constructed and the plant formulation is re-formulated for ACC systems considering \eqref{revied_Hinf_ACC}. The revised augmented plant for ACC systems is illustrated in Fig. \ref{P_new2_fig} and given by:
\begin{eqnarray}\label{P_new2_eq}
P =
\left[
  \begin{array}{c|c}
  W_{S} & -W_{S} G H \\
    0   & -W_{U} \\
    0   & W_{T} G \\
    0   & G \\
    \hline
    I   & -G H \\
  \end{array}
  \right]
\end{eqnarray}

Since the CACC structure is different from the ACC structure, the augmented plant is reformulated for the multi-objective $H_{\infty}$ control method. The revised augmented plant for the CACC systems is illustrated in Fig. \ref{P_new_CACC_fig} and given by:
\begin{figure}[h!]
\begin{center}
  \includegraphics[width=3.4 in]{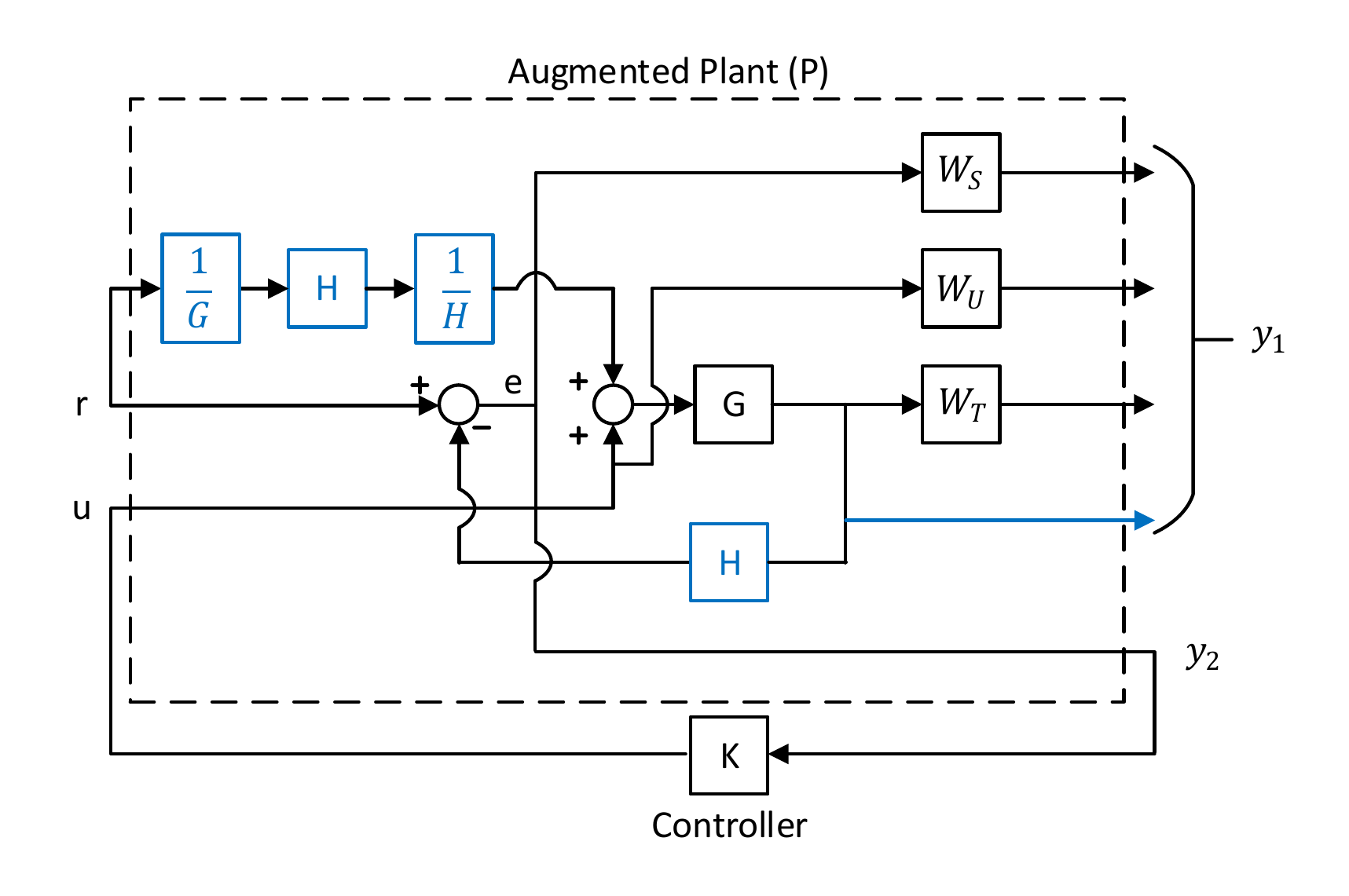}\\
  \caption{New Augmented Plant for CACC systems}\label{P_new_CACC_fig}
   \end{center}
\end{figure}

\begin{eqnarray}\label{P_new_CACC_eq}
P =
\left[
  \begin{array}{c|c}
  W_{S} (1-D) & -W_{S} G H \\
    0   & -W_{U} \\
    W_{T} D / H   & W_{T} G \\
     D / H   &  G \\
   \hline
        I-D   & -G H \\
  \end{array}
  \right]
\end{eqnarray}

\section{Simulation Studies}\label{SimulationStudies}

The first step in the controller design is to define the control objectives clearly. In ACC and CACC systems, the objective is to drive a following car at a desired relative distance with respect to its preceding one while the string stability is satisfied and the closed-loop uncertain system is robustly stable.
  
The parameters for the controller design procedure are selected as follows \cite{SSCACC}: the time-constant $\tau$ and the time-delay $\phi$ of the system model in \eqref{longitudinaltf} are respectively equal to $0.1$ and $0.2$ second. The communication delay between vehicles in a platoon $\theta$ in \eqref{Dtimedelay} is equal to $0.15$ second. The time-headway parameter $h$ of the velocity dependent spacing policy in \eqref{H_tf} is respectively chosen as $1$ and $0.5$ second for ACC and CACC systems. It is assumed that inter-vehicle distances are initially larger than the desired relative distance determined by the spacing policy. In other words, the relative spacing error between vehicles is positive.

The sampling time depends on beaconing, communication protocol, vehicle speeds and number of vehicles in a platoon and is generally set to $10$ Hz \cite{Ploeg2011jmt}. The controller design is realized in discrete-time domain and the $4^{th}$ order Pade approximation is used to model the time-delays by using rational models. It is to be noted that high-order Pade approximations result in transfer functions with clustered poles so that pole configurations tend to be very sensitive to perturbations. Therefore, a low-order Pade approximation is preferred in this study as inspired by \cite{Xing2017}.

The control problem for (C)ACC systems is formulated as follows:
\begin{eqnarray}\label{multi_Hinf_CACC}
\displaystyle\min\limits_{K \in \mathcal{K}, \Gamma_{S}, \Gamma_{T} } && \Gamma_{S} + \Gamma_{T} \nonumber \\
\textrm{subject to} && \| W_{S} S \|_{\infty} \leq \Gamma_{S} \nonumber \\
                && \| W_{T} T \|_{\infty} \leq \Gamma_{T} \nonumber \\
              && \| T \|_{\infty} \leq   M_{T} =1 
\end{eqnarray}
If $\Gamma_{S}$ and $\Gamma_{T}$ $<$ $1$, the sufficient conditions are satisfied for the controller design. There are two closed-loop subsystems to be tried to minimize their $H_{\infty}$-norms separately. It is to be noted that the input sensitivity $U$ is not evaluated throughout controller design. The weighting matrices  $W_{S}$ and $W_{T}$ are selected for the performance and robustness as follows:
\begin{eqnarray}\label{Weightingmatrices}
W_{S} & = & 0.035\frac{z^{2}}{(z -0.99)(z -0.99)} \\
W_{T} & = & 0.3\frac{(z -0.99)(z -0.99)}{z^{2}}
\end{eqnarray}
$W_{S}$ and $W_{T}$ are respectively the performance and robustness weighting functions to limit the magnitudes of the sensitivity and complementary sensitivity functions. The main aim in robust control problem is to reduce the effect of disturbance on output so that the sensitivity function S and the complementary sensitivity function T are to be reduced. However, they cannot be kept small over the whole frequency range due to the $S + T = I$ constraint. Therefore, some trade-off between the sensitivity and complementary sensitivity functions must be done. In general, reference signals and disturbances occur at low frequencies while modeling errors and sensor noise occur at high frequencies. The sensitivity function S and the complementary sensitivity function T must be kept small at low frequencies and high frequencies, respectively. As a result, low gain for the sensitivity function at low frequencies is for good tracking while low gain for the complementary sensitivity function at high frequencies makes system insensitive to disturbances. Moreover, high gain for the sensitivity function at high frequencies limits overshoot. 

\begin{remark} 
It is to be noted that if there is no solution for the control formulation in \eqref{multi_Hinf_CACC}, it is necessary to change the weighting matrices $W_{S}$ and $W_{T}$, or the headway time $h_{i}$ in \eqref{H_tf} to obtain a solution.  
\end{remark}

\begin{figure}[t!]
\begin{center}
  \includegraphics[width=3.3in]{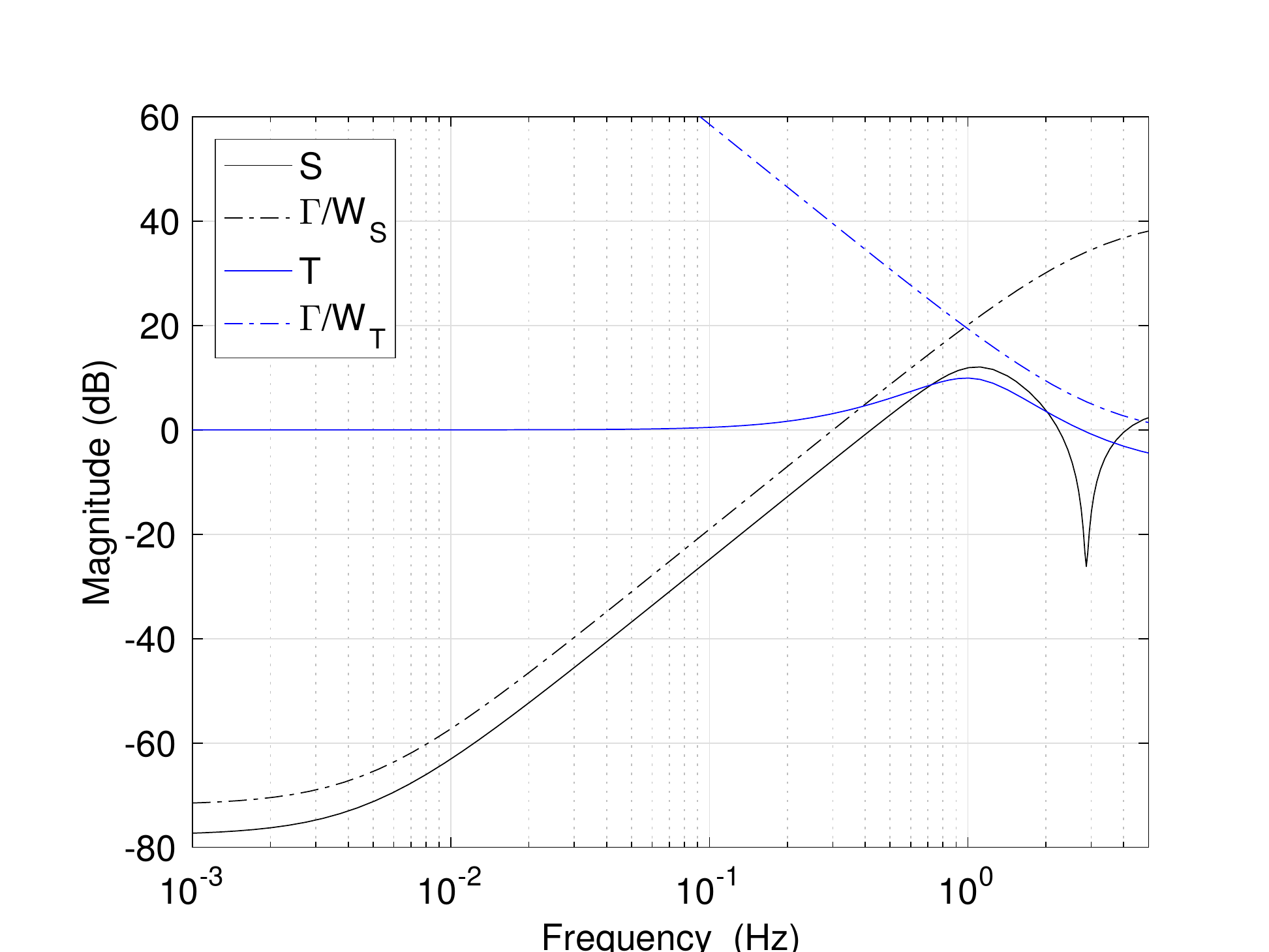}\\
  \caption{The frequency domain function response of the traditional $H_{\infty}$ control for ACC}\label{ACC_ST}
   \end{center}
\end{figure}

The frequency domain function response of the traditonal $H_{\infty}$ control structure for the ACC systems is illustrated in Fig. \ref{ACC_ST}. As can be seen, the complementary sensitivity function $\| T \|_{\infty}$ is larger than $0$ dB. This shows that the string stability of the system could not be achieved for the time headway $h$, which is equal to $1$ second.

\begin{figure}[t!]
\begin{center}
  \includegraphics[width=3.3in]{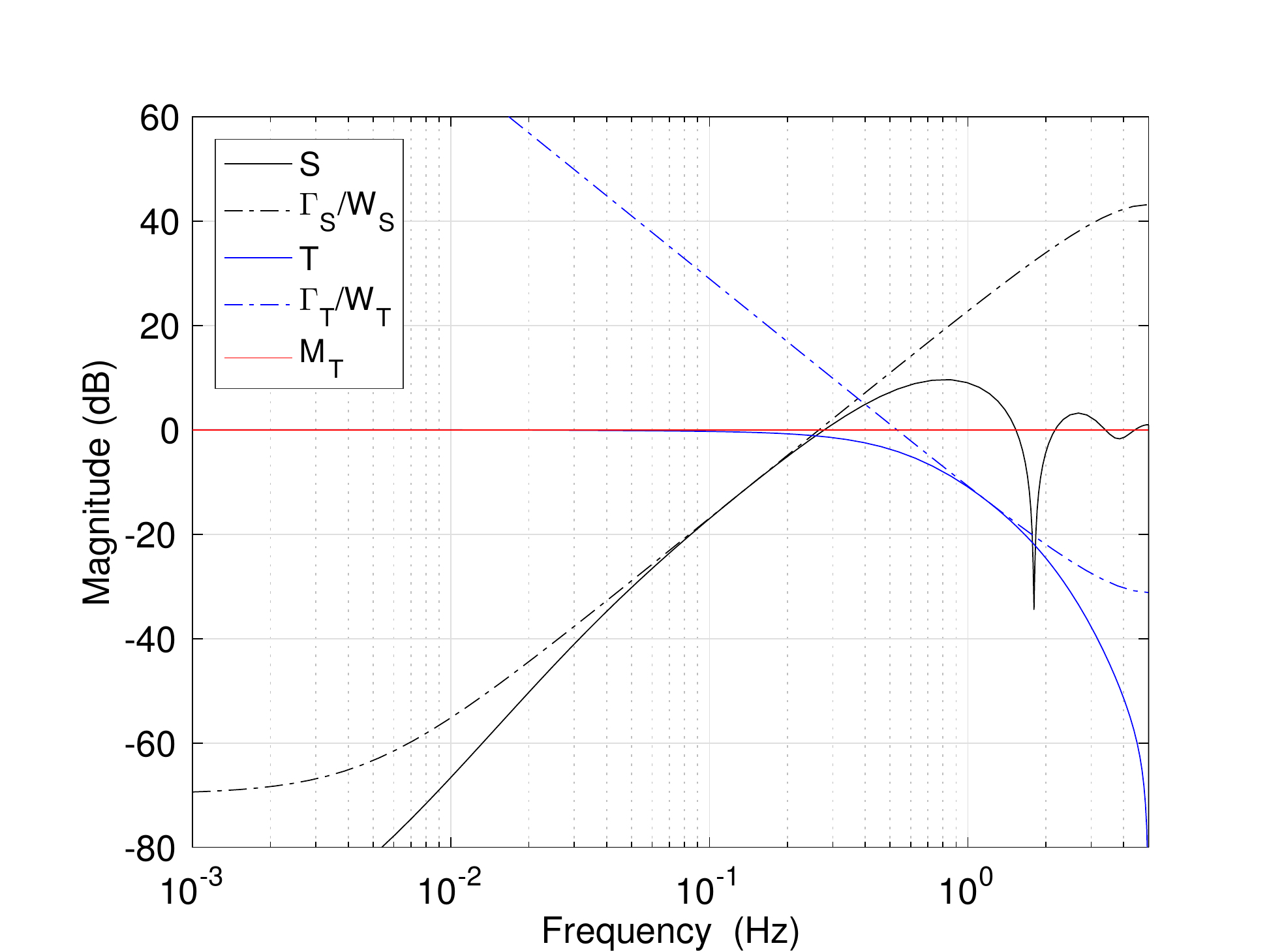}\\
  \caption{The frequency domain function response of the multi-objective $H_{\infty}$ control for ACC}\label{ACC_ST_revised}
   \end{center}
\end{figure}

The frequency domain function response of the multi-objective $H_{\infty}$ control structure is illustrated in Fig. \ref{ACC_ST_revised}. As can be seen, the complementary sensitivity function $\| T \|_{\infty}$ is equal and smaller than $0$ dB. Thus, it is concluded that the string stability of the system is achieved for the time headway $h$, which is equal to $1$ second. In the previous studies as shown in Table \ref{timeheadways}, it was found $2.6$ and $3.16$ seconds for the same model with the same actuator delay \cite{SSCACC,Ploeg2011}. 

\begin{figure}[t!]
\begin{center}
  \includegraphics[width=3.3in]{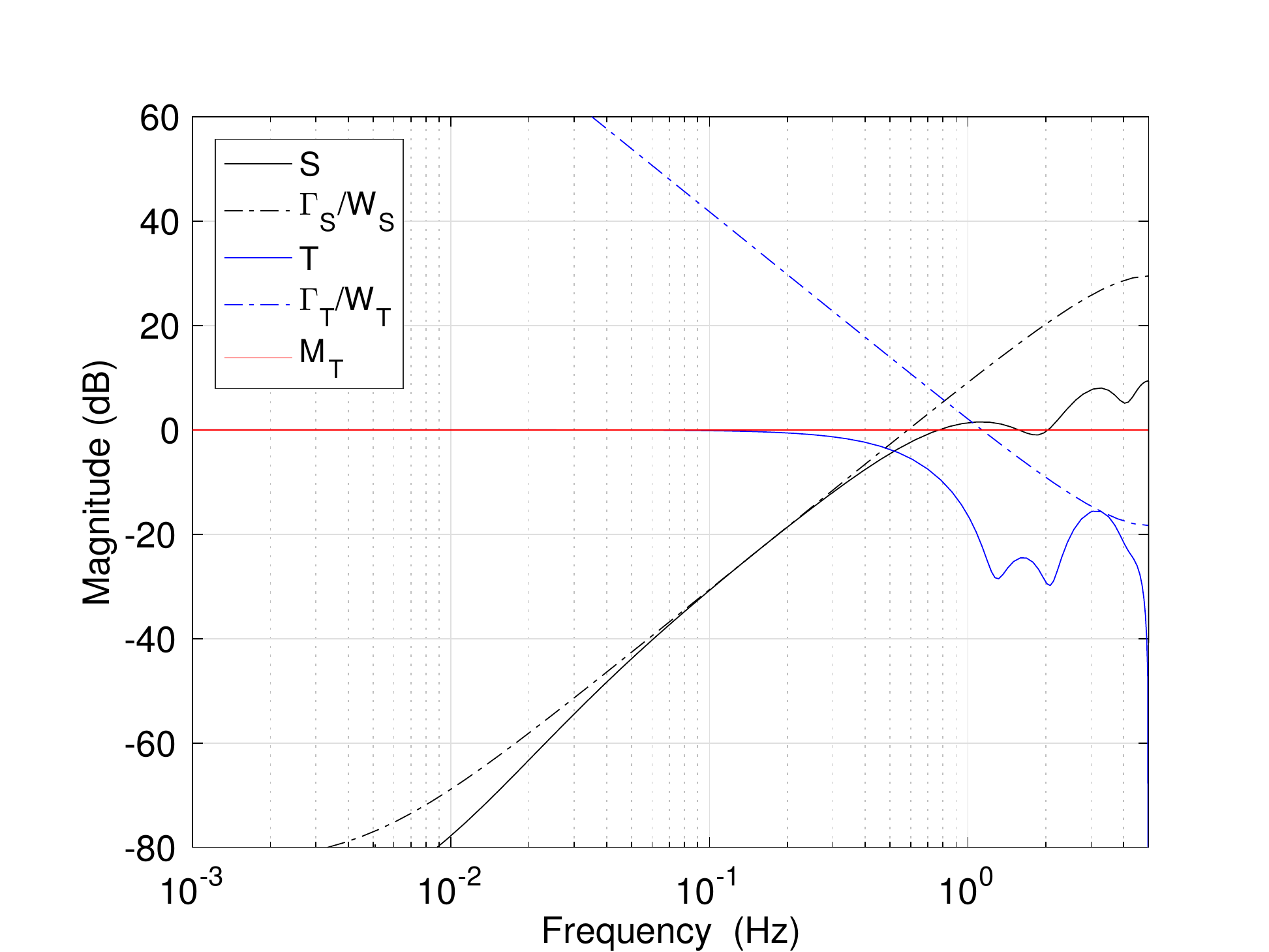}\\
  \caption{The frequency domain function response of the multi-objective $H_{\infty}$ control for CACC}\label{CACC_ST_revised}
   \end{center}
\end{figure}

The frequency domain function response of the multi-objective $H_{\infty}$ control structure for CACC systems is shown in Fig. \ref{CACC_ST_revised}. As can be seen, the complementary sensitivity function $\| T \|_{\infty}$ is equal and smaller than $0$ dB. Thus, it is concluded that the string stability of the system is achieved for the time headway $h$, which is equal to $0.5$ second. In the previous studies as shown in Table \ref{timeheadways}, it was found $0.7$ and $0.8$ second for the same model with the same actuator and communication delays \cite{Ploeg2011,Ploeg2014Lp}. Moreover, it was found $0.8$ second while the communication delay is equal to $60$ ms \cite{SSCACC}. It can be concluded that the multi-objective $H_{\infty}$ control method has a great potential when compared to previous studies. 

Since feedback controller reduces disturbances with frequencies less than crossover frequency, it is desirable to move the crossover frequency to high frequencies by choosing the weighting matrix  $W_{S}$.  As can be seen Figs. \ref{ACC_ST_revised}-\ref{CACC_ST_revised}, the crossover frequencies are respectively around $0.27$ Hz and $0.8$ Hz for ACC and CACC systems. This shows that CACC systems have ability to reduce disturbances better than ACC systems. 


\begin{table}[h!]
\centering
\caption{Time-headway for ACC and CACC systems}\label{timeheadways}
\begin{tabular}{lcc}
  \hline
   &  ACC & CACC  \\
   &  (s) & (s)  \\
    \hline
   Proposed Multi-$H_{\infty}$ & 1 &  0.5  \\
    In \cite{Ploeg2011} & 3.16 & 0.7 \\
    In \cite{SSCACC} & 2.6 &  0.8  \\
  In \cite{Ploeg2014Lp} & - &  0.7 \\
  \hline
\end{tabular}
\end{table}

\begin{figure*}[h!]
\centering
\subfigure[ ]{
\includegraphics[width=3.2in]{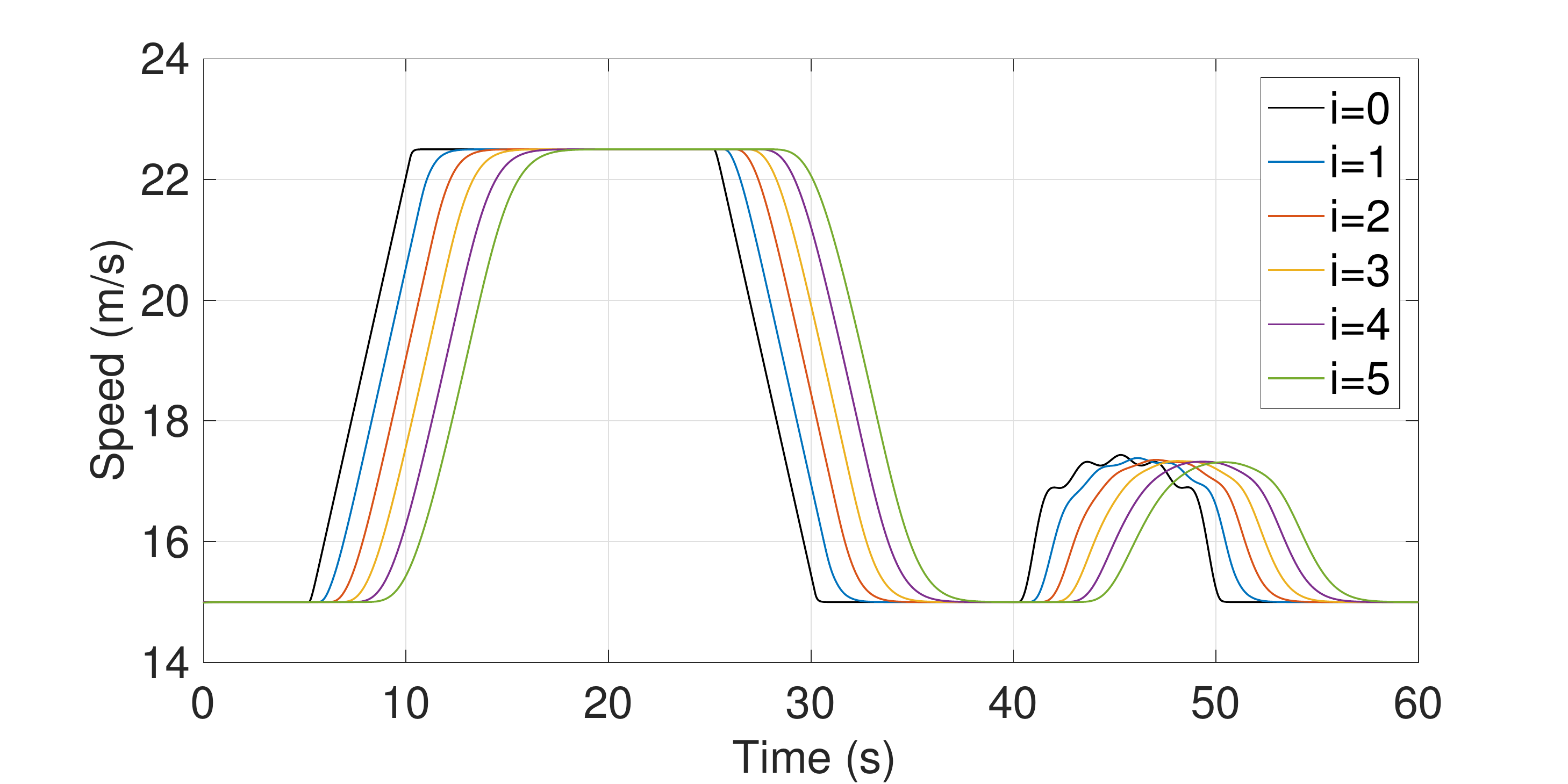}
\label{ACC_speed}
}
\subfigure[ ]{
\includegraphics[width=3.2in]{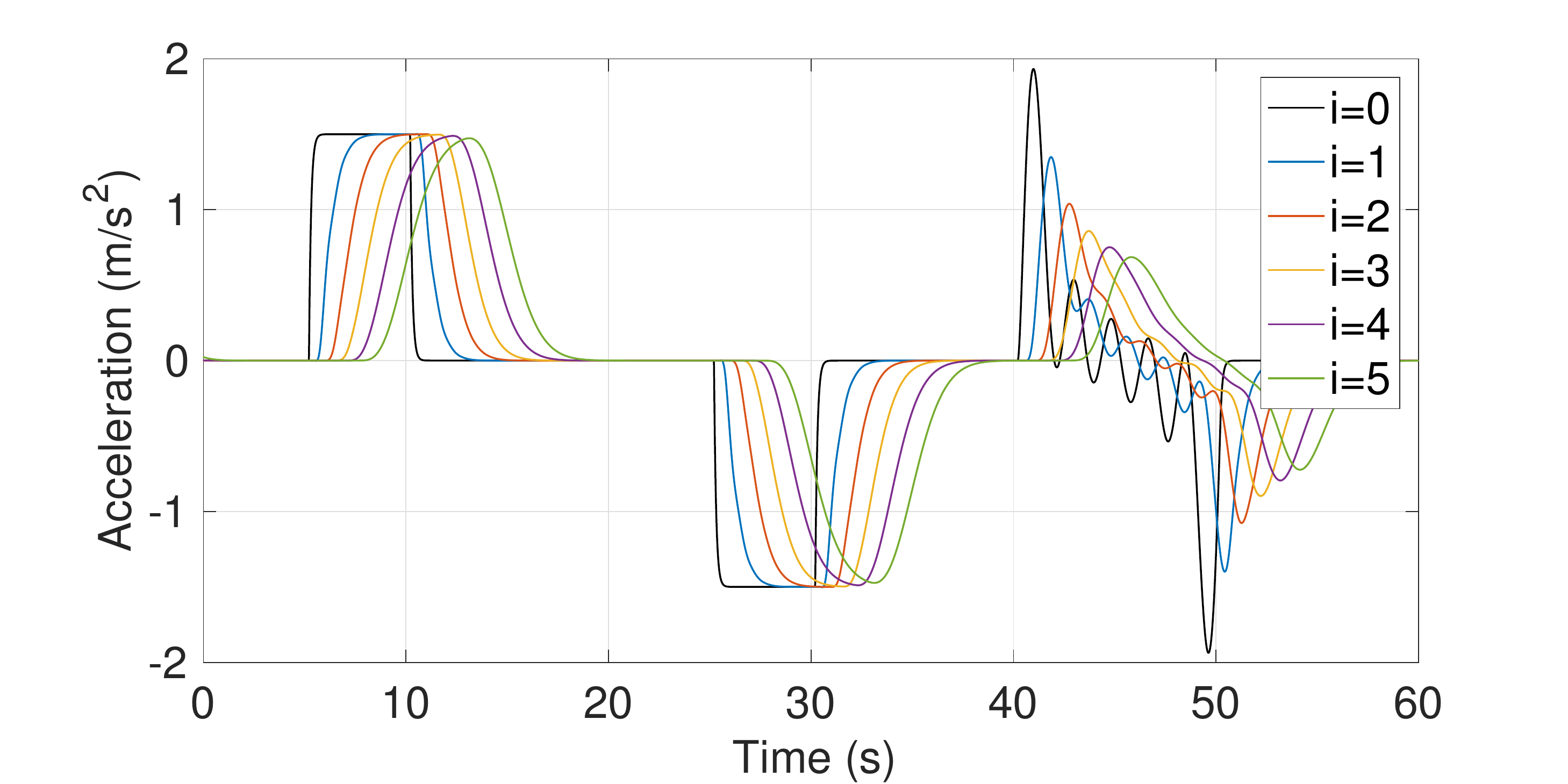}
\label{CACC_speed}
}
\subfigure[ ]{
\includegraphics[width=3.2in]{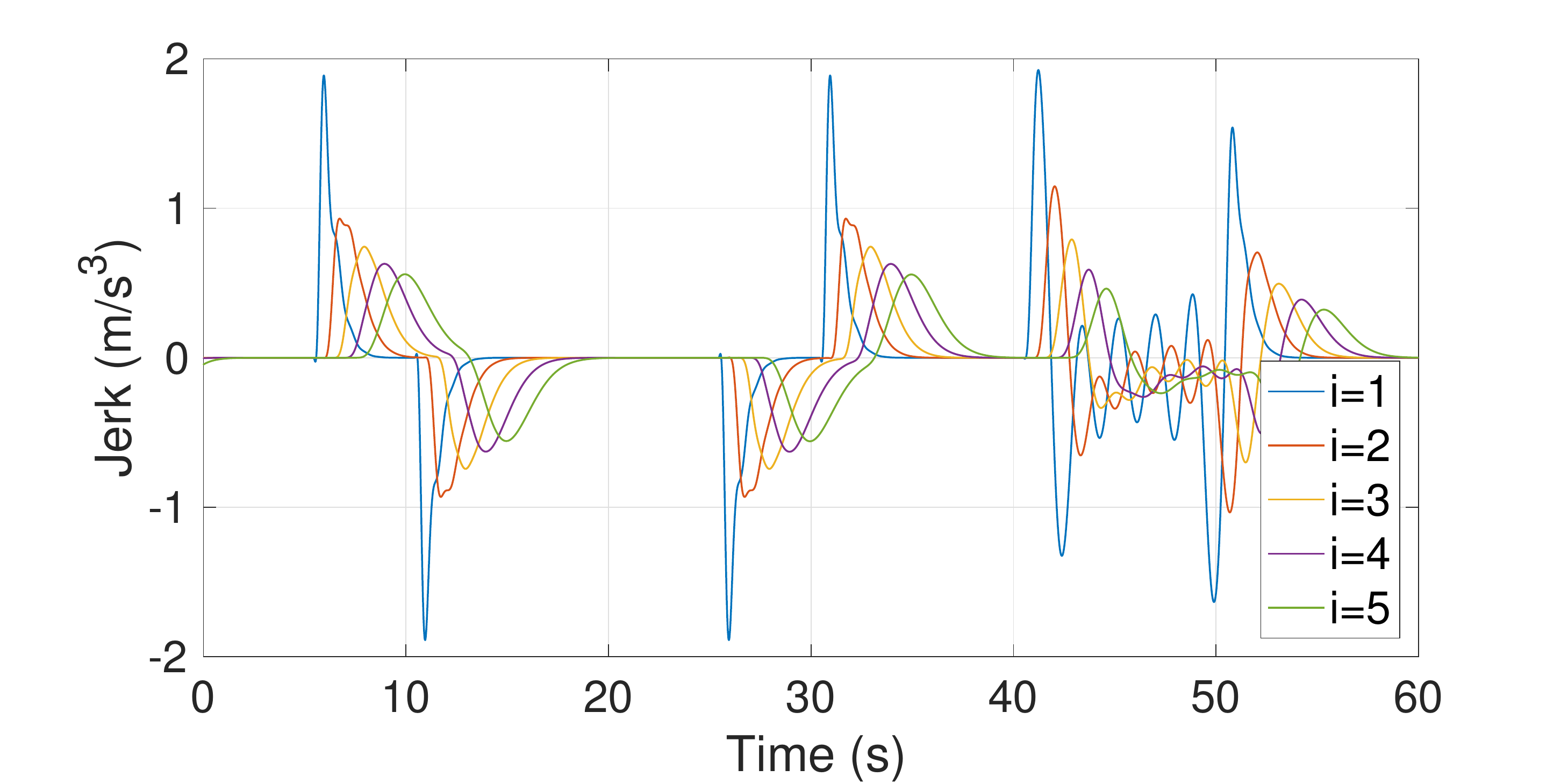}
\label{CACC_speed}
}
\subfigure[ ]{
\includegraphics[width=3.2in]{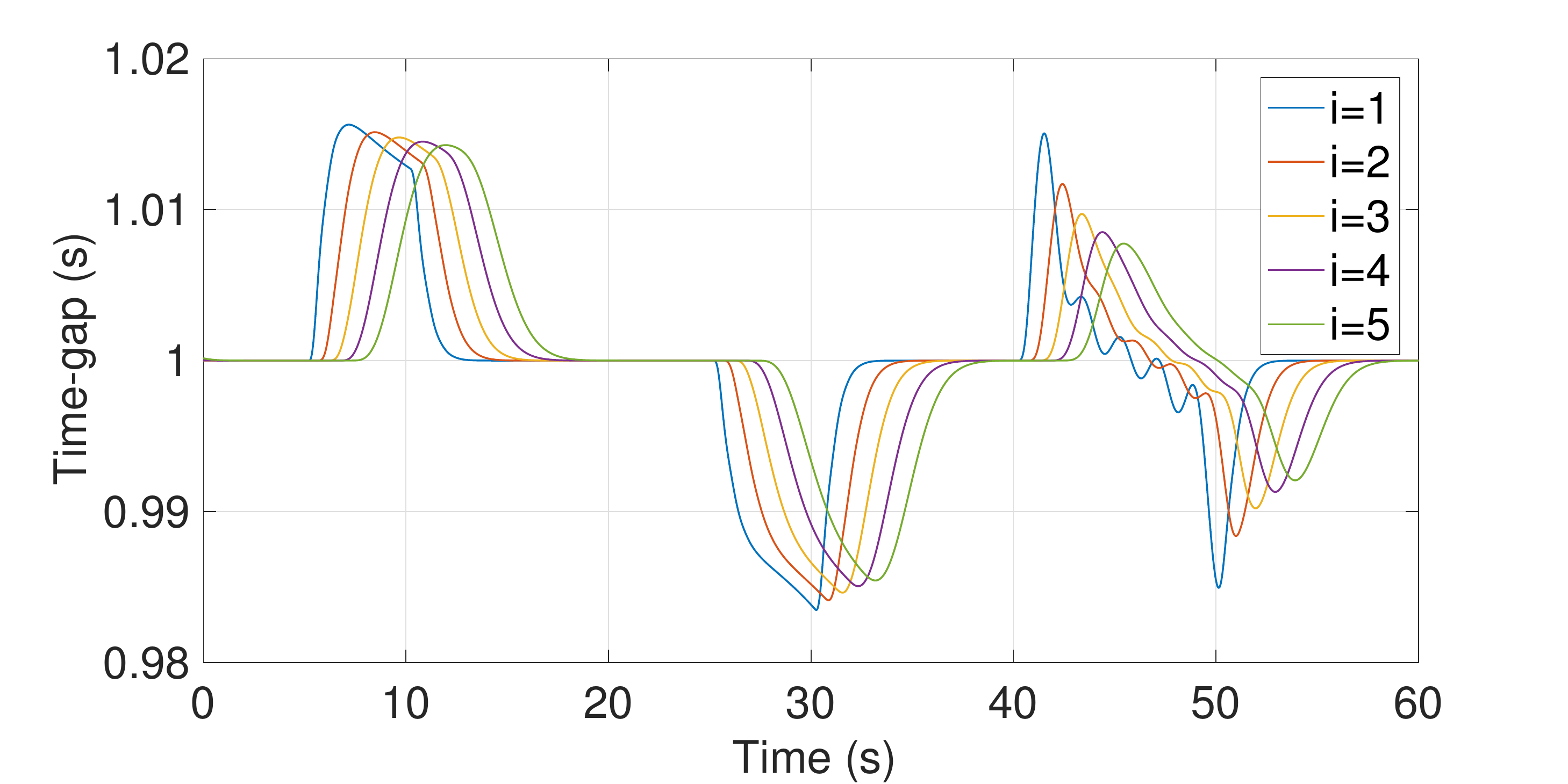}
\label{CACC_speed}
}
\subfigure[ ]{
\includegraphics[width=3.2in]{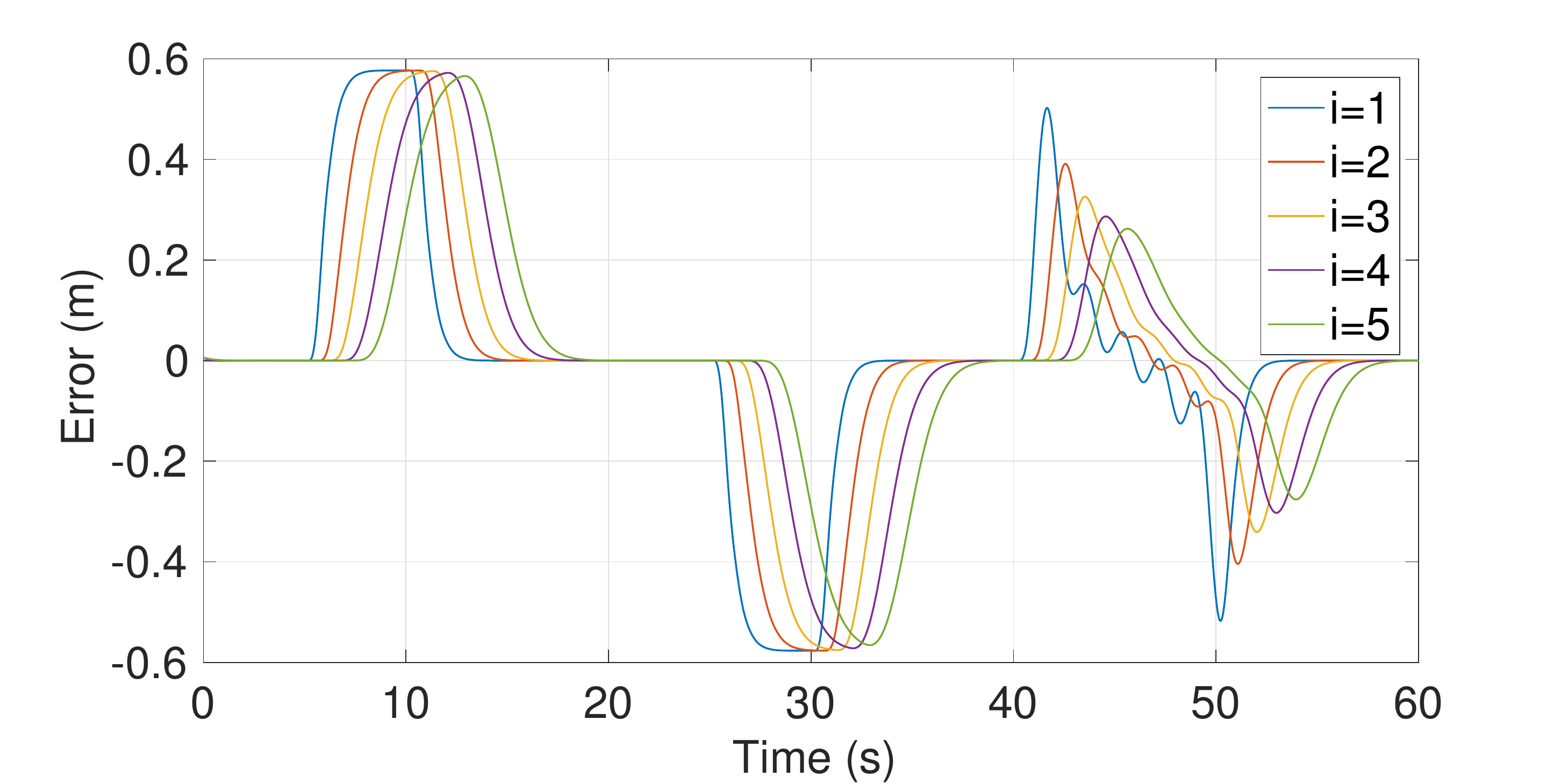}
\label{CACC_speed}
}
\subfigure[ ]{
\includegraphics[width=3.2in]{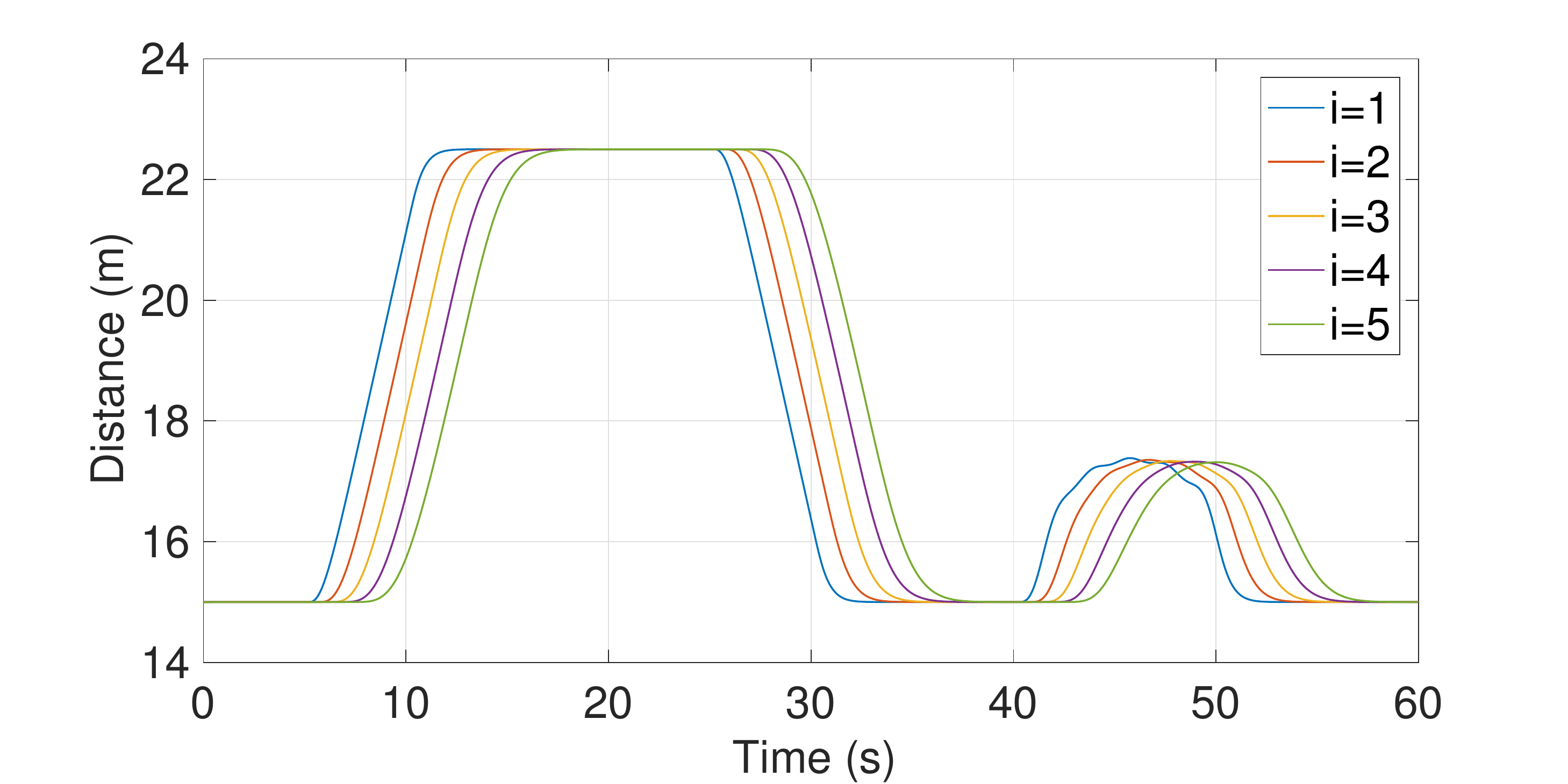}
\label{CACC_speed}
}
\caption[Optional caption for list of figures]{ Time responses of (a) speed and (b) acceleration (i =0, 1, 2, 3, 4 , 5) and of (c) jerk, (d) time-gap, (e) distance error and (e) distance i = 1, 2, 3, 4, 5) for ACC case. }
\label{sim_ACC}
\end{figure*}

\begin{figure*}[h!]
\centering
\subfigure[ ]{
\includegraphics[width=3.2in]{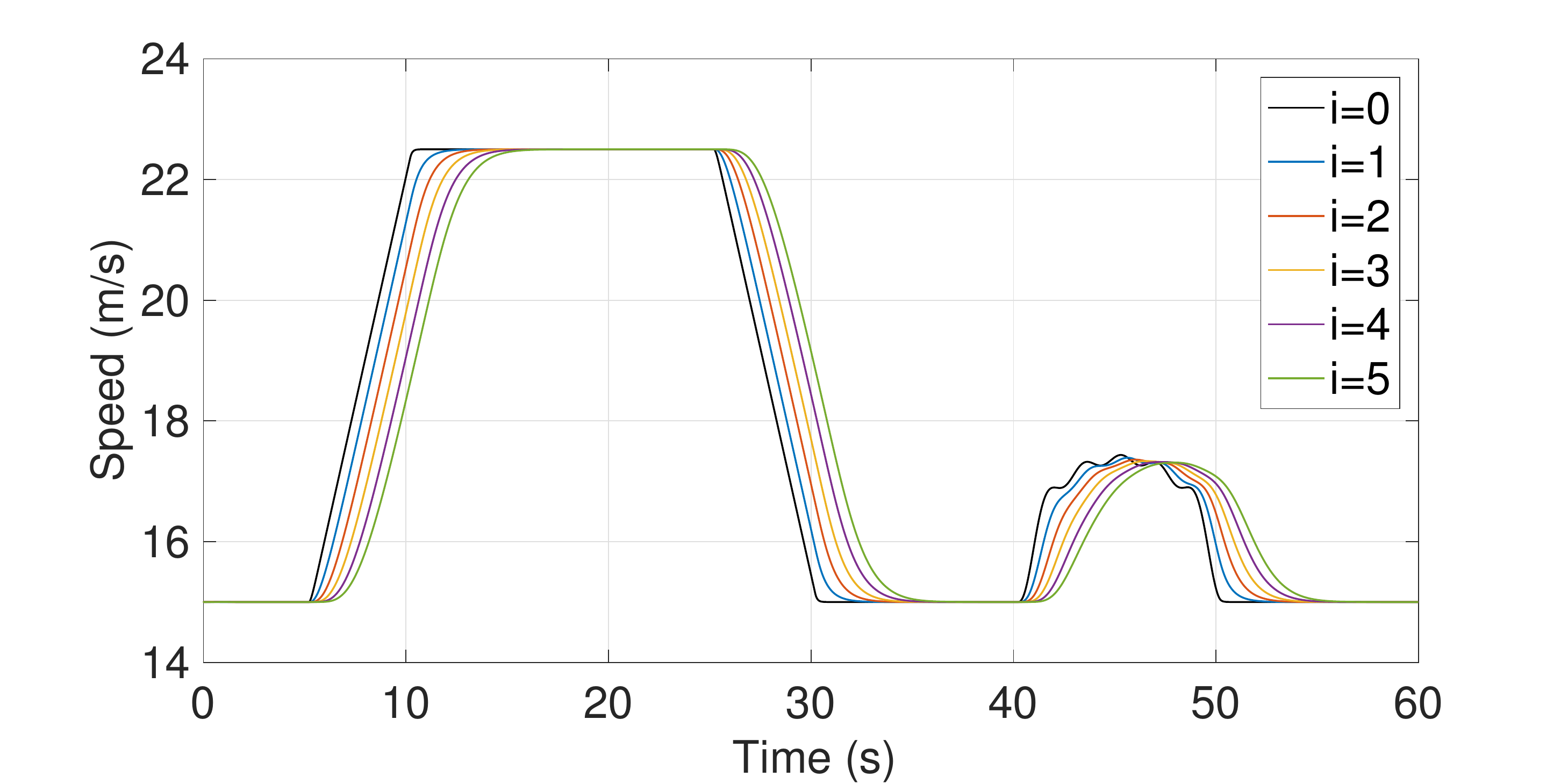}
\label{ACC_speed}
}
\subfigure[ ]{
\includegraphics[width=3.2in]{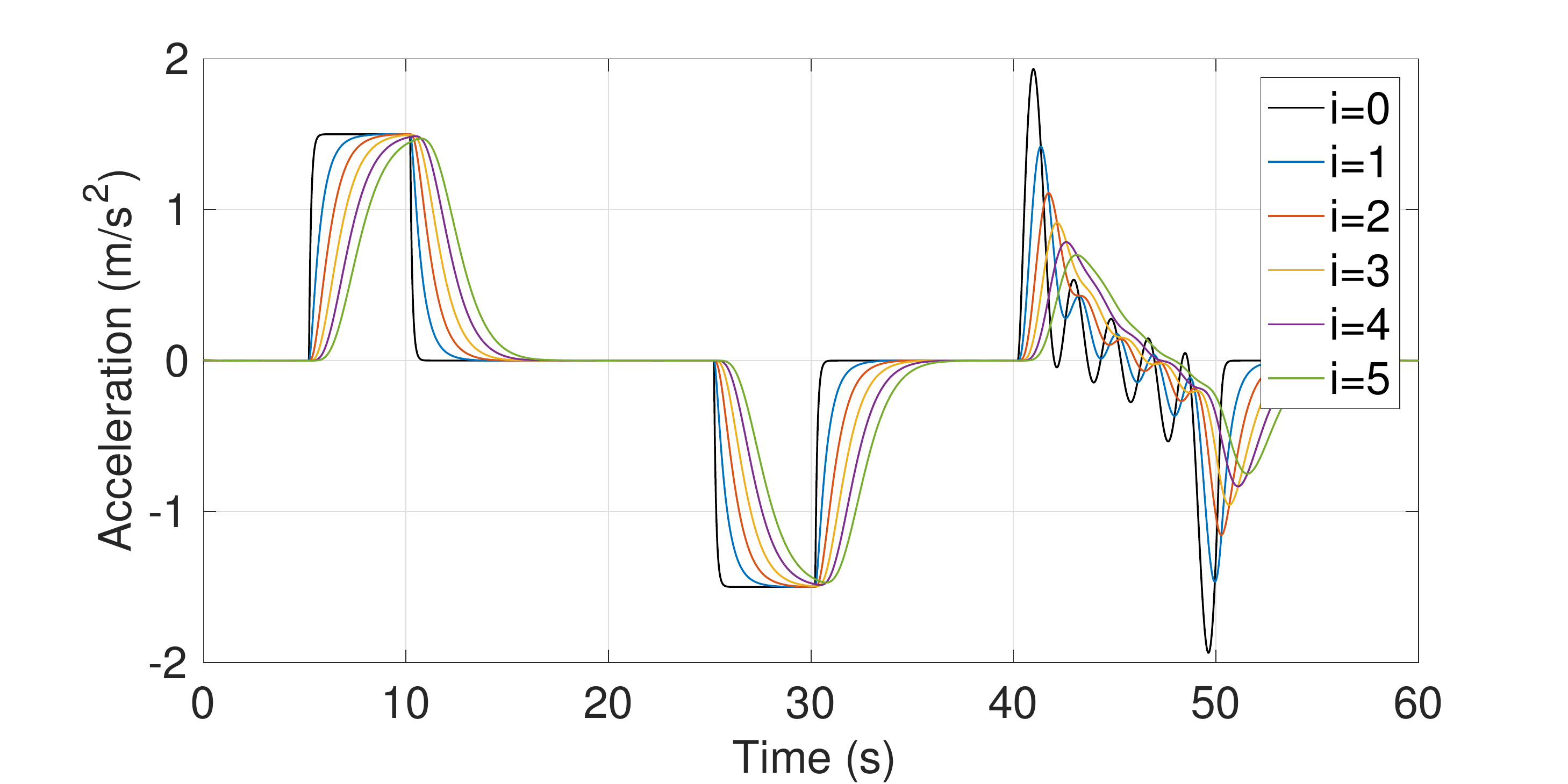}
\label{CACC_speed}
}
\subfigure[ ]{
\includegraphics[width=3.2in]{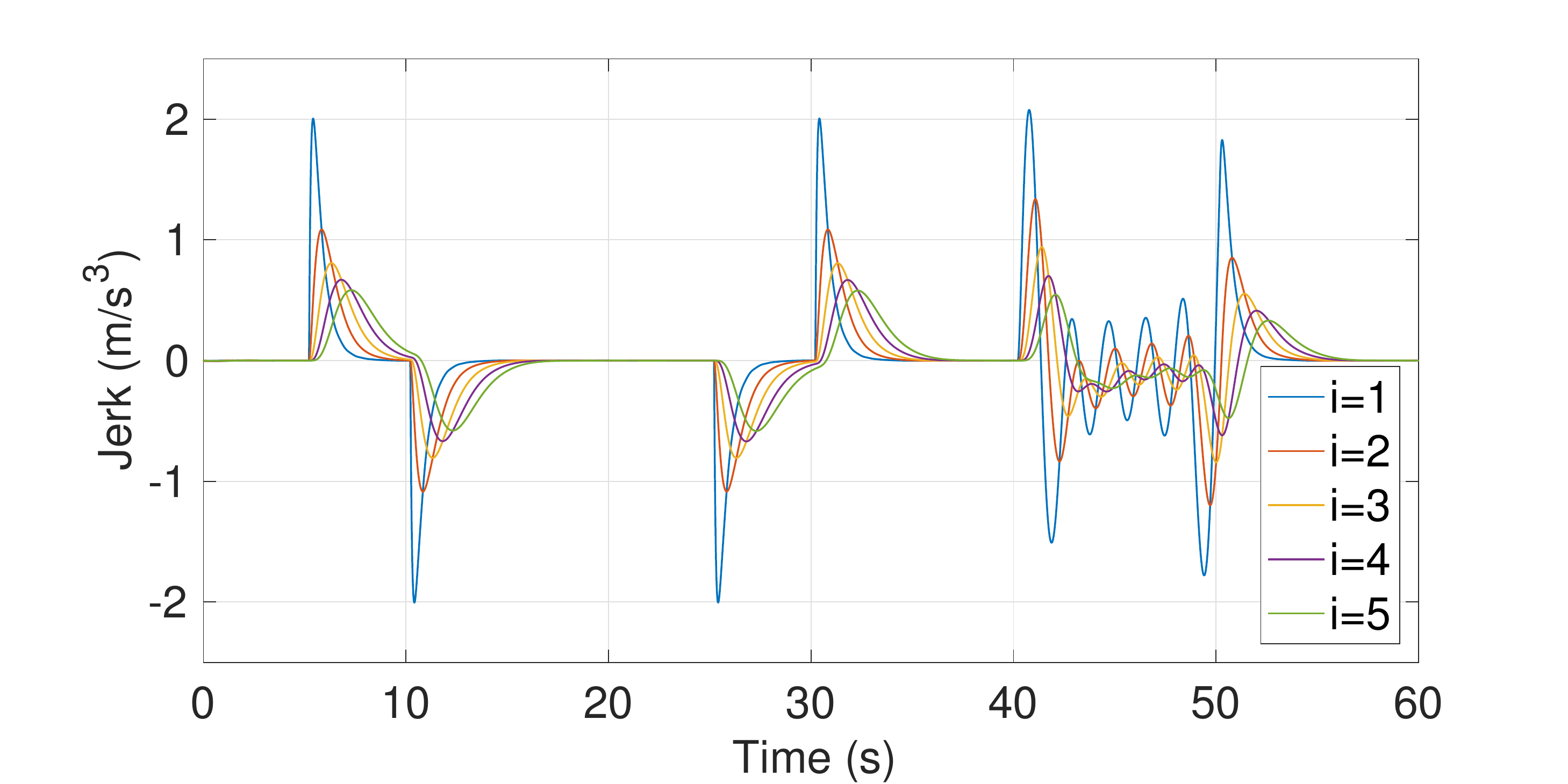}
\label{CACC_speed}
}
\subfigure[ ]{
\includegraphics[width=3.2in]{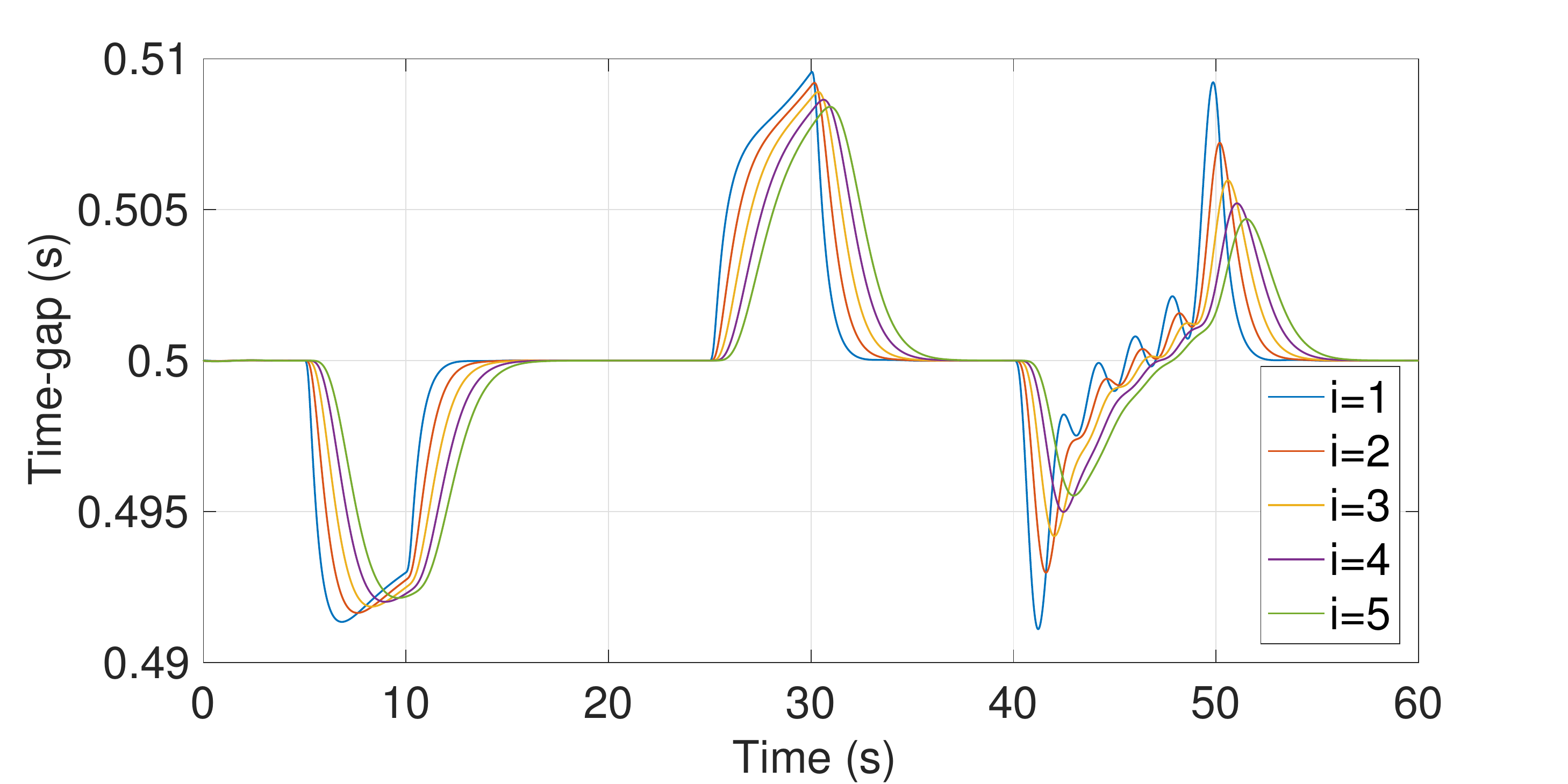}
\label{CACC_speed}
}
\subfigure[ ]{
\includegraphics[width=3.2in]{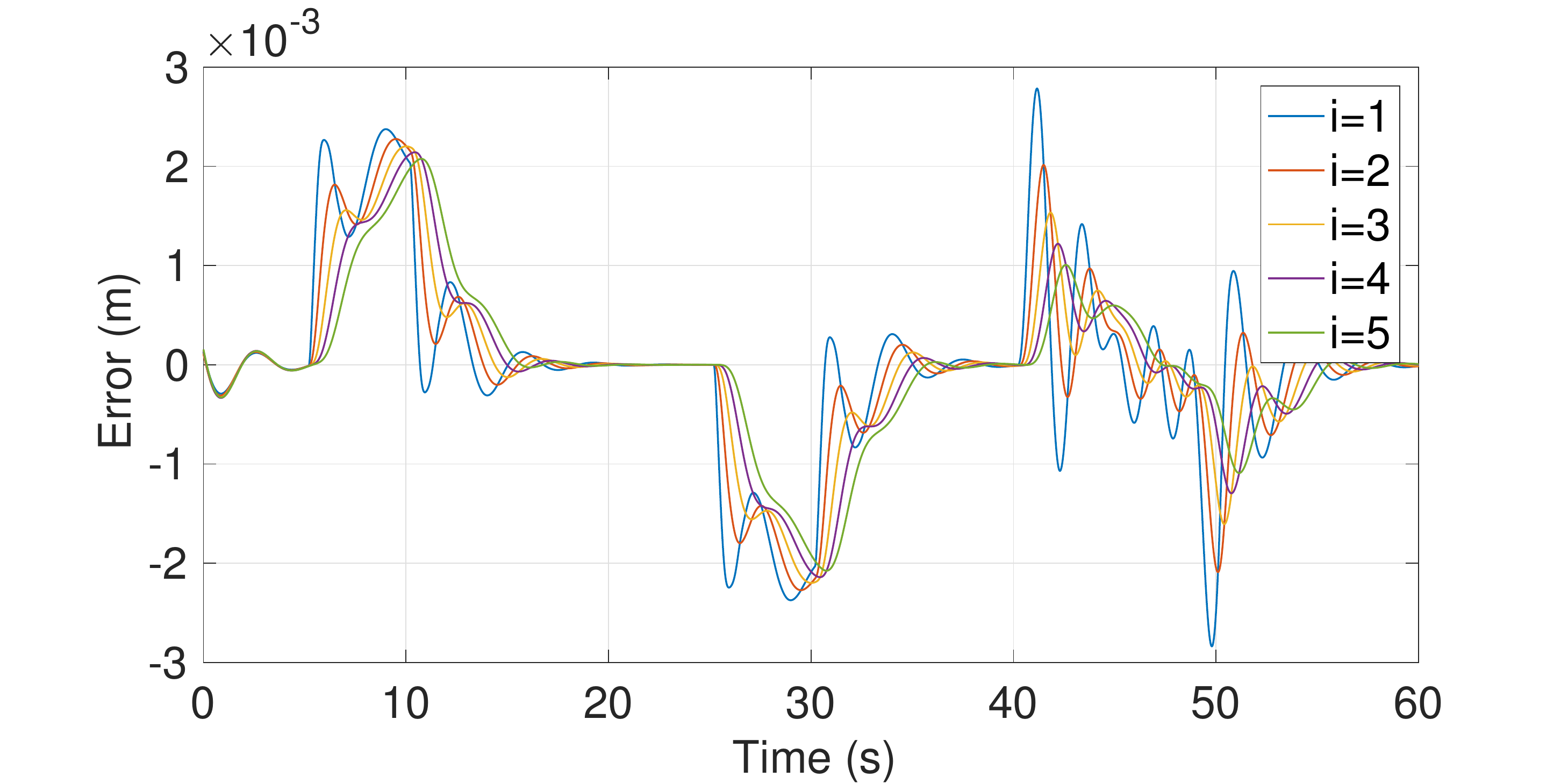}
\label{CACC_speed}
}
\subfigure[ ]{
\includegraphics[width=3.2in]{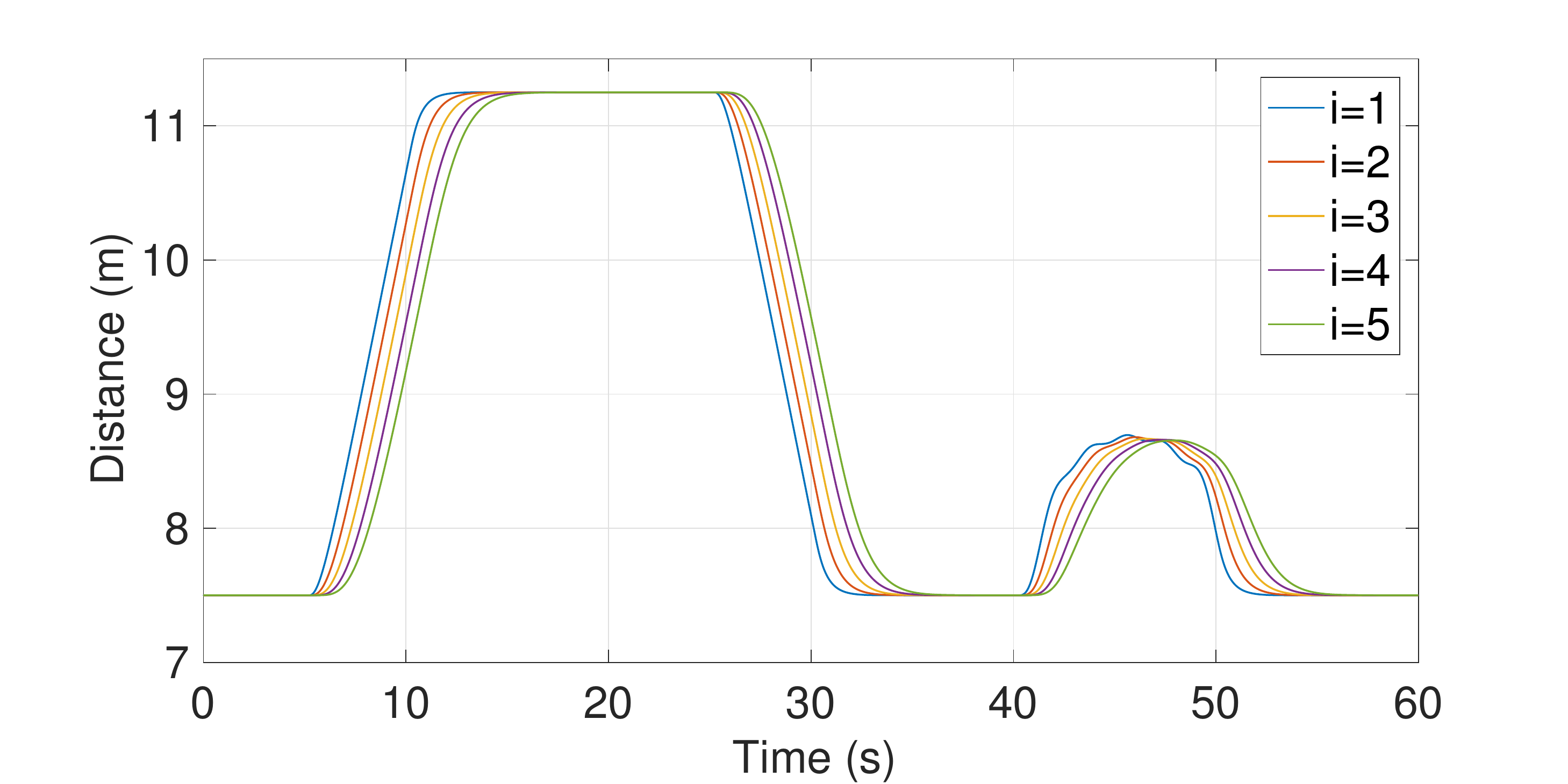}
\label{CACC_speed}
}
\caption[Optional caption for list of figures]{ Time responses of (a) speed and (b) acceleration (i = 0, 1, 2, 3, 4 , 5) and of (c) jerk, (d) time-gap, (e) distance error and (e) distance i = 1, 2, 3, 4, 5) for CACC case. }
\label{sim_CACC}
\end{figure*}

The test trajectory is defined by the desired acceleration $u_{i-1}(t)$ of the lead vehicle, and consists of step and multisine signals as follows:
\begin{equation}\label{eq_aref}
  u_{i-1}(t) = \Bigg \{
    \begin{array}{rl}
    5 \leq t < 10  & d= 1.5 \\
   25 \leq t < 30 & d= -1.5 \\
   40 \leq t < 50 & d= 0.5 \sum_{k=1} ^{5} \sin{(0.1k t)} 
 \end{array} 
\end{equation}
The time responses of one leading and five following vehicles for ACC and CACC cases are shown in Figs. \ref{sim_ACC}-\ref{sim_CACC}. The acceleration and velocity responses show a decreasing amplitude along the string while the desired acceleration $u_{i-1}(t)$ of the lead vehicle is a multisine signal. This indicates string stable behavior which is also confirmed by the absence of overshoot in the acceleration and speed responses while the desired acceleration $u_{i-1}(t)$ of the lead vehicle is a step signal. Moreover, the distance error of each preceding vehicle is larger than the error of each following vehicle. It represents the string stability of the platoon due to the fact that the string stability condition in time-domain is defined as follows:
\begin{equation}
\max{e_{i}}<\max{e_{i-1}}, \quad 1 \leq i \leq m
\end{equation}
where $e_{i}$ is the error of the vehicle $i$.  Furthermore, the distance and time-hap responses reflect the velocity dependent spacing policy. Furthermore, it is to be noted that the following vehicles in CACC systems start accelerating before the following vehicles in ACC systems. This demonstrates that CACC may also be effective at traffic lights.

\begin{figure}[h!]
\begin{center}
  \includegraphics[width=3.3in]{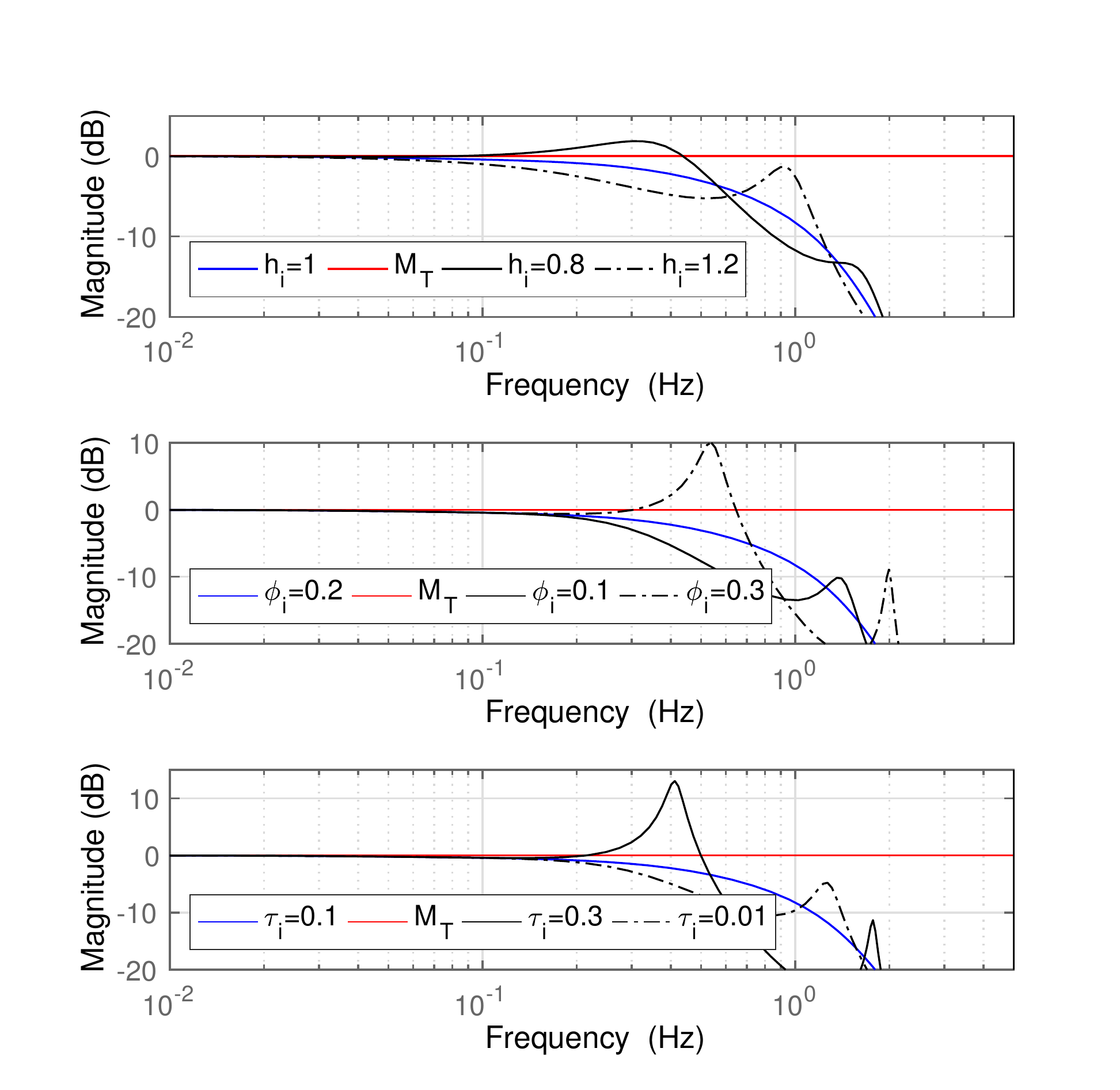}\\
  \caption{The frequency domain function responses of ACC for different time-gaps, time-delays, time-lags and communication delays.}\label{ACC_fd_different}
   \end{center}
\end{figure}

\begin{figure}[h!]
\begin{center}
  \includegraphics[width=3.3in]{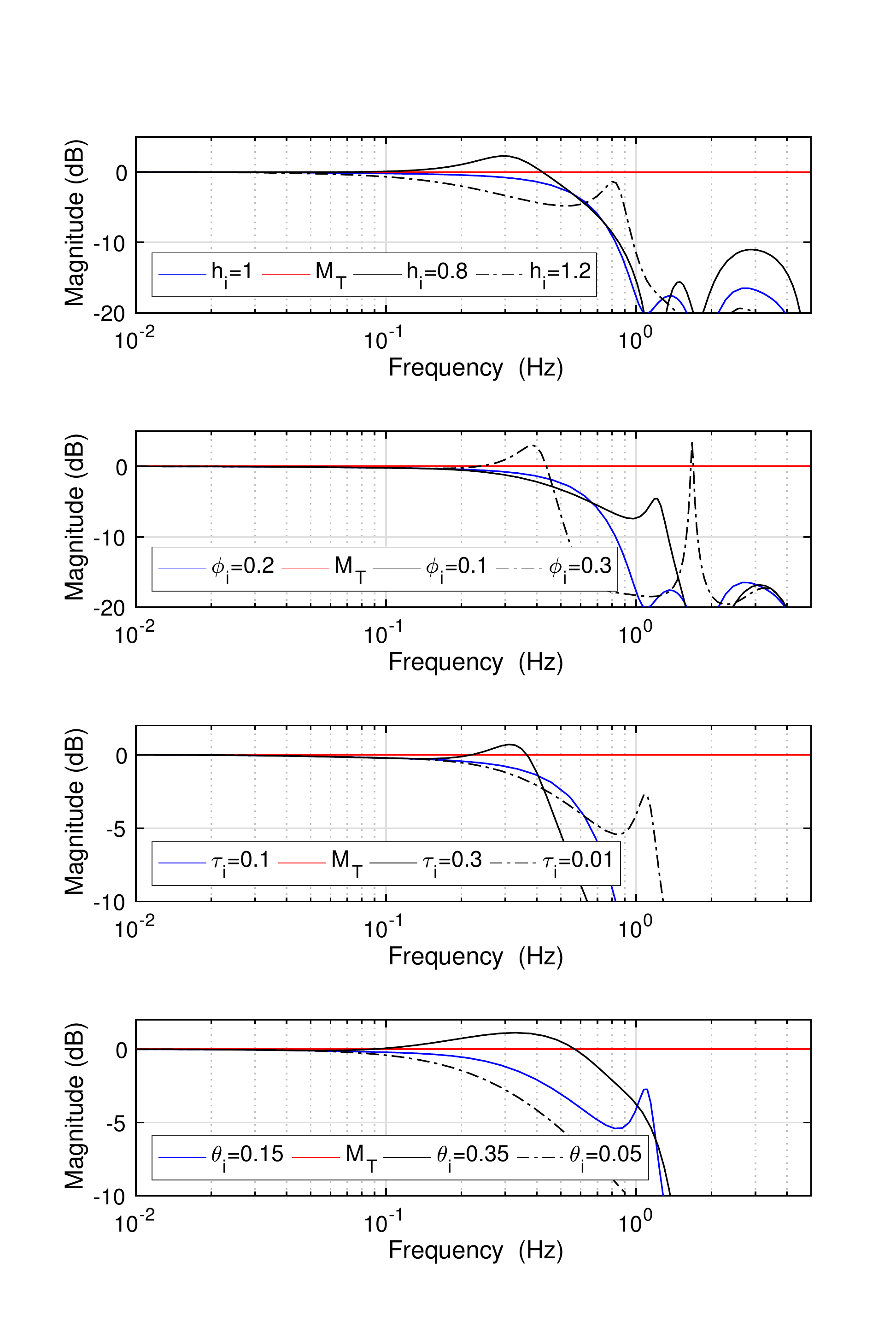}\\
  \caption{The frequency domain function responses of CACC for different time-gaps, time-delays, time-lags and communication delays.}\label{CACC_fd_different}
   \end{center}
\end{figure}

The frequency domain responses of different time-gaps, time-delays, time-lags and communication delays for ACC and CACC systems are shown in Figs. \ref{ACC_fd_different}-\ref{CACC_fd_different}. As can be seen from these figures, smaller time-gaps, and larger time-delays and time-lag result in string instability. It is concluded that the controller obtained in \eqref{multi_Hinf_CACC} satisfies the string stability if the time headway is larger than the selected time headway, e.g. respectively $h_{i}>1$ and $h_{i}>0.5$ for ACC and CACC, and time-delays and time-lags are smaller than the selected time-delays and time-lags, e.g. $\phi_{i}<0.2$, $\theta_{i}<0.15$, $\tau_{i}<0.1$.


\section{Conclusions}\label{Conclusions}

A multi-objective $H_{\infty}$ control method has been elaborated for ACC and CACC systems in the frequency-domain. The simulation studies show that the proposed control algorithm results in smaller inter-vehicle distances when compared to the traditional one. Moreover, it is able to provide not only the system stability but also the string stability. Since the proposed control algorithm is model-based control approach, the identification of system parameters in the longitudinal model is very crucial. Implementations of the controllers are feasible in real-time due to the fact that no online solution is required and distributed control structure is used. Experimental results are more than welcome to verify the validity of the simulation results for future research. 

\bibliography{ref_Hinf}

\begin{thebibliography}{10}
\providecommand{\url}[1]{#1}
\csname url@samestyle\endcsname
\providecommand{\newblock}{\relax}
\providecommand{\bibinfo}[2]{#2}
\providecommand{\BIBentrySTDinterwordspacing}{\spaceskip=0pt\relax}
\providecommand{\BIBentryALTinterwordstretchfactor}{4}
\providecommand{\BIBentryALTinterwordspacing}{\spaceskip=\fontdimen2\font plus
\BIBentryALTinterwordstretchfactor\fontdimen3\font minus
  \fontdimen4\font\relax}
\providecommand{\BIBforeignlanguage}[2]{{%
\expandafter\ifx\csname l@#1\endcsname\relax
\typeout{** WARNING: IEEEtran.bst: No hyphenation pattern has been}%
\typeout{** loaded for the language `#1'. Using the pattern for}%
\typeout{** the default language instead.}%
\else
\language=\csname l@#1\endcsname
\fi
#2}}
\providecommand{\BIBdecl}{\relax}
\BIBdecl

\bibitem{Ashley2016}
A.~B. McDonald, D.~V. McGehee, S.~T. Chrysler, N.~M. Askelson, L.~S. Angell,
  and B.~D. Seppelt, ``National survey identifying gaps in consumer knowledge
  of advanced vehicle safety systems,'' \emph{Transportation Research Record:
  Journal of the Transportation Research Board}, vol. 2559, pp. 1--6, 2016.

\bibitem{Vahidi2003}
A.~Vahidi and A.~Eskandarian, ``Research advances in intelligent collision
  avoidance and adaptive cruise control,'' \emph{Intelligent Transportation
  Systems, IEEE Transactions on}, vol.~4, no.~3, pp. 143--153, 2003.

\bibitem{Rajamani2002}
R.~Rajamani and C.~Zhu, ``Semi-autonomous adaptive cruise control systems,''
  \emph{Vehicular Technology, IEEE Transactions on}, vol.~51, no.~5, pp.
  1186--1192, Sep 2002.

\bibitem{Ploeg2014Lp}
J.~Ploeg, N.~van~de Wouw, and H.~Nijmeijer, ``Lp string stability of cascaded
  systems: Application to vehicle platooning,'' \emph{Control Systems
  Technology, IEEE Transactions on}, vol.~22, no.~2, pp. 786--793, 2014.

\bibitem{Naus2010}
G.~Naus, R.~Vugts, J.~Ploeg, R.~Van De~Molengraft, and M.~Steinbuch,
  ``Cooperative adaptive cruise control, design and experiments,'' in
  \emph{American Control Conference (ACC), 2010}, June 2010, pp. 6145--6150.

\bibitem{Ploeg2011}
J.~Ploeg, B.~Scheepers, E.~van Nunen, N.~van~de Wouw, and H.~Nijmeijer,
  ``Design and experimental evaluation of cooperative adaptive cruise
  control,'' in \emph{Intelligent Transportation Systems (ITSC), 2011 14th
  International IEEE Conference on}, Oct 2011, pp. 260--265.

\bibitem{Guvenc2012}
L.~Guuvenc, I.~Uygan, K.~Kahraman, R.~Karaahmetoglu, I.~Altay, M.~Sentuurk,
  M.~Emirler, A.~Karci, B.~Guuvenc, E.~Altug, M.~Turan, O.~Tas, E.~Bozkurt,
  U.~Ozguner, K.~Redmill, A.~Kurt, and B.~Efendioglu, ``Cooperative adaptive
  cruise control implementation of team mekar at the grand cooperative driving
  challenge,'' \emph{Intelligent Transportation Systems, IEEE Transactions on},
  vol.~13, no.~3, pp. 1062--1074, 2012.

\bibitem{Milanes2014}
V.~Milanes, S.~Shladover, J.~Spring, C.~Nowakowski, H.~Kawazoe, and
  M.~Nakamura, ``Cooperative adaptive cruise control in real traffic
  situations,'' \emph{Intelligent Transportation Systems, IEEE Transactions
  on}, vol.~15, no.~1, pp. 296--305, 2014.

\bibitem{Fanping2010}
F.~Bu, H.-S. Tan, and J.~Huang, ``Design and field testing of a cooperative
  adaptive cruise control system,'' in \emph{American Control Conference (ACC),
  2010}, June 2010, pp. 4616--4621.

\bibitem{Kianfar2012}
R.~Kianfar, B.~Augusto, A.~Ebadighajari, U.~Hakeem, J.~Nilsson, A.~Raza,
  R.~Tabar, N.~Irukulapati, C.~Englund, P.~Falcone, S.~Papanastasiou,
  L.~Svensson, and H.~Wymeersch, ``Design and experimental validation of a
  cooperative driving system in the grand cooperative driving challenge,''
  \emph{Intelligent Transportation Systems, IEEE Transactions on}, vol.~13,
  no.~3, pp. 994--1007, Sept 2012.

\bibitem{Desjardins2011}
C.~Desjardins and B.~Chaib-draa, ``Cooperative adaptive cruise control: A
  reinforcement learning approach,'' \emph{Intelligent Transportation Systems,
  IEEE Transactions on}, vol.~12, no.~4, Dec 2011.

\bibitem{Liang1999}
C.-Y. Liang and H.~Peng, ``Optimal adaptive cruise control with guaranteed
  string stability,'' \emph{Vehicle System Dynamics}, vol.~32, no. 4-5, pp.
  313--330, 1999.

\bibitem{erkanDiNMPC}
E.~Kayacan, E.~Kayacan, H.~Ramon, and W.~Saeys, ``Distributed nonlinear model
  predictive control of an autonomous tractor–trailer system,''
  \emph{Mechatronics}, vol.~24, no.~8, pp. 926 -- 933, 2014.

\bibitem{erkanDeNMPC}
------, ``Robust tube-based decentralized nonlinear model predictive control of
  an autonomous tractor-trailer system,'' \emph{IEEE/ASME Transactions on
  Mechatronics}, vol.~20, no.~1, pp. 447--456, Feb 2015.

\bibitem{Dunbar2012}
W.~Dunbar and D.~Caveney, ``Distributed receding horizon control of vehicle
  platoons: Stability and string stability,'' \emph{Automatic Control, IEEE
  Transactions on}, vol.~57, no.~3, pp. 620--633, March 2012.

\bibitem{Liu2015}
B.~Liu and A.~El~Kamel, ``V2x-based decentralized cooperative adaptive cruise
  control in the vicinity of intersections,'' \emph{Intelligent Transportation
  Systems, IEEE Transactions on}, vol.~PP, no.~99, pp. 1--1, 2015.

\bibitem{Hao2013}
\BIBentryALTinterwordspacing
H.~Hao and P.~Barooah, ``Stability and robustness of large platoons of vehicles
  with double-integrator models and nearest neighbor interaction,''
  \emph{International Journal of Robust and Nonlinear Control}, vol.~23,
  no.~18, pp. 2097--2122, 2013. [Online]. Available:
  \url{http://dx.doi.org/10.1002/rnc.2872}
\BIBentrySTDinterwordspacing

\bibitem{Ploeg2014Synthesis}
J.~Ploeg, D.~P. Shukla, N.~van~de Wouw, and H.~Nijmeijer, ``Controller
  synthesis for string stability of vehicle platoons,'' \emph{IEEE Transactions
  on Intelligent Transportation Systems}, vol.~15, no.~2, pp. 854--865, 2014.

\bibitem{oncu2014}
S.~Oncu, J.~Ploeg, N.~van~de Wouw, and H.~Nijmeijer, ``Cooperative adaptive
  cruise control: Network-aware analysis of string stability,''
  \emph{Intelligent Transportation Systems, IEEE Transactions on}, vol.~15,
  no.~4, pp. 1527--1537, Aug 2014.

\bibitem{pirani2016graph}
M.~Pirani, E.~Hashemi, J.~W. Simpson-Porco, B.~Fidan, and A.~Khajepour, ``Graph
  theoretic approach to the robustness of k-nearest neighbor vehicle
  platoons,'' \emph{IEEE Transactions on Intelligent Transportation Systems},
  pp. 1--7, 2017.

\bibitem{swaroop1994}
D.~Swaropp, J.~hedrick, C.~C. Chien, and P.~Ioannou, ``A comparision of spacing
  and headway control laws for automatically controlled vehicles,''
  \emph{Vehicle System Dynamics}, vol.~23, no.~1, pp. 597--625, 1994.

\bibitem{Wang2004}
J.~Wang and R.~Rajamani, ``Should adaptive cruise-control systems be designed
  to maintain a constant time gap between vehicles?'' \emph{IEEE Transactions
  on Vehicular Technology}, vol.~53, no.~5, pp. 1480--1490, Sept 2004.

\bibitem{Jing2005}
J.~Zhou and H.~Peng, ``Range policy of adaptive cruise control vehicles for
  improved flow stability and string stability,'' \emph{Intelligent
  Transportation Systems, IEEE Transactions on}, vol.~6, no.~2, pp. 229--237,
  2005.

\bibitem{sheikholeslam1992}
S.~Sheikholeslam and C.~A. Desoer, ``A system level study of the longitudinal
  control of a platoon of vehicles,'' \emph{Journal of dynamic systems,
  measurement, and control}, vol. 114, no.~2, pp. 286--292, 1992.

\bibitem{Santhanakrishnan2003}
K.~Santhanakrishnan and R.~Rajamani, ``On spacing policies for highway vehicle
  automation,'' \emph{Intelligent Transportation Systems, IEEE Transactions
  on}, vol.~4, no.~4, pp. 198--204, Dec 2003.

\bibitem{swaroop1999}
D.~Swaroop and J.~Hedrick, ``Constant spacing strategies for platooning in
  automated highway systems,'' \emph{Journal of dynamic systems, measurement,
  and control}, vol. 121, no.~3, pp. 462--470, 1999.

\bibitem{Liu2001}
X.~Liu, A.~Goldsmith, S.~Mahal, and J.~Hedrick, ``Effects of communication
  delay on string stability in vehicle platoons,'' in \emph{Intelligent
  Transportation Systems, 2001. Proceedings. 2001 IEEE}, 2001, pp. 625--630.

\bibitem{Seiler2004}
P.~Seiler, A.~Pant, and K.~Hedrick, ``Disturbance propagation in vehicle
  strings,'' \emph{Automatic Control, IEEE Transactions on}, vol.~49, no.~10,
  pp. 1835--1842, Oct 2004.

\bibitem{SSCACC}
G.~Naus, R.~Vugts, J.~Ploeg, M.~van~de Molengraft, and M.~Steinbuch,
  ``String-stable cacc design and experimental validation: A frequency-domain
  approach,'' \emph{Vehicular Technology, IEEE Transactions on}, vol.~59,
  no.~9, pp. 4268--4279, Nov 2010.

\bibitem{Xiao2011}
L.~Xiao and F.~Gao, ``Practical string stability of platoon of adaptive cruise
  control vehicles,'' \emph{IEEE Transactions on Intelligent Transportation
  Systems}, vol.~12, no.~4, pp. 1184--1194, Dec 2011.

\bibitem{Sheikholeslam1993}
S.~Sheikholeslam and C.~A. Desoer, ``Longitudinal control of a platoon of
  vehicles with no communication of lead vehicle information: a system level
  study,'' \emph{IEEE Transactions on Vehicular Technology}, vol.~42, no.~4,
  pp. 546--554, Nov 1993.

\bibitem{Stankovic2000}
S.~S. Stankovic, M.~J. Stanojevic, and D.~D. Siljak, ``Decentralized
  overlapping control of a platoon of vehicles,'' \emph{IEEE Transactions on
  Control Systems Technology}, vol.~8, no.~5, pp. 816--832, Sep 2000.

\bibitem{Caveney2010}
D.~Caveney, ``Cooperative vehicular safety applications,'' \emph{Control
  Systems, IEEE}, vol.~30, no.~4, pp. 38--53, Aug 2010.

\bibitem{Wischhof2005}
L.~Wischhof, A.~Ebner, and H.~Rohling, ``Information dissemination in
  self-organizing intervehicle networks,'' \emph{Intelligent Transportation
  Systems, IEEE Transactions on}, vol.~6, no.~1, pp. 90--101, March 2005.

\bibitem{Vinel2015}
A.~Vinel, L.~Lan, and N.~Lyamin, ``Vehicle-to-vehicle communication in
  c-acc/platooning scenarios,'' \emph{Communications Magazine, IEEE}, vol.~53,
  no.~8, pp. 192--197, August 2015.

\bibitem{Dey2015}
K.~Dey, L.~Yan, X.~Wang, Y.~Wang, H.~Shen, M.~Chowdhury, L.~Yu, C.~Qiu, and
  V.~Soundararaj, ``A review of communication, driver characteristics, and
  controls aspects of cooperative adaptive cruise control (cacc),''
  \emph{Intelligent Transportation Systems, IEEE Transactions on}, vol.~PP,
  no.~99, pp. 1--19, 2015.

\bibitem{Zhou198885}
K.~Zhou and P.~P. Khargonekar, ``An algebraic riccati equation approach to
  ${H}_{\infty}$ optimization,'' \emph{Systems \& Control Letters}, vol.~11,
  no.~2, pp. 85 -- 91, 1988.

\bibitem{Gahinet1994}
P.~Gahinet and P.~Apkarian, ``A linear matrix inequality approach to
  ${H}_{\infty}$ control,'' \emph{International Journal of Robust and Nonlinear
  Control}, vol.~4, no.~4, pp. 421--448, 1994.

\bibitem{scherer1997}
C.~Scherer, P.~Gahinet, and M.~Chilali, ``Multiobjective output-feedback
  control via {LMI} optimization,'' \emph{Automatic Control, IEEE Transactions
  on}, vol.~42, no.~7, pp. 896--911, Jul 1997.

\bibitem{Oliveira2002}
M.~C.~D. Oliveira, J.~C. Geromel, and J.~Bernussou, ``Extended {H} 2 and {H}
  norm characterizations and controller parametrizations for discrete-time
  systems,'' \emph{International Journal of Control}, vol.~75, no.~9, pp.
  666--679, 2002.

\bibitem{Ploeg2011jmt}
\BIBentryALTinterwordspacing
J.~Ploeg, A.~F.~A. Serrarens, and G.~J. Heijenk, ``Connect {\&} drive: design
  and evaluation of cooperative adaptive cruise control for congestion
  reduction,'' \emph{Journal of Modern Transportation}, vol.~19, no.~3, pp.
  207--213, 2011. [Online]. Available:
  \url{http://dx.doi.org/10.1007/BF03325760}
\BIBentrySTDinterwordspacing

\bibitem{Xing2017}
H.~Xing, J.~Ploeg, and H.~Nijmeijer, ``Pade; approximation of delays in
  cooperative acc based on string stability requirements,'' \emph{IEEE
  Transactions on Intelligent Vehicles}, vol.~PP, no.~99, pp. 1--1, 2017.

\end{thebibliography}
\bibliographystyle{IEEEtran}

\begin{IEEEbiography}[{\includegraphics[width=1in,height=1.25in,clip,keepaspectratio]{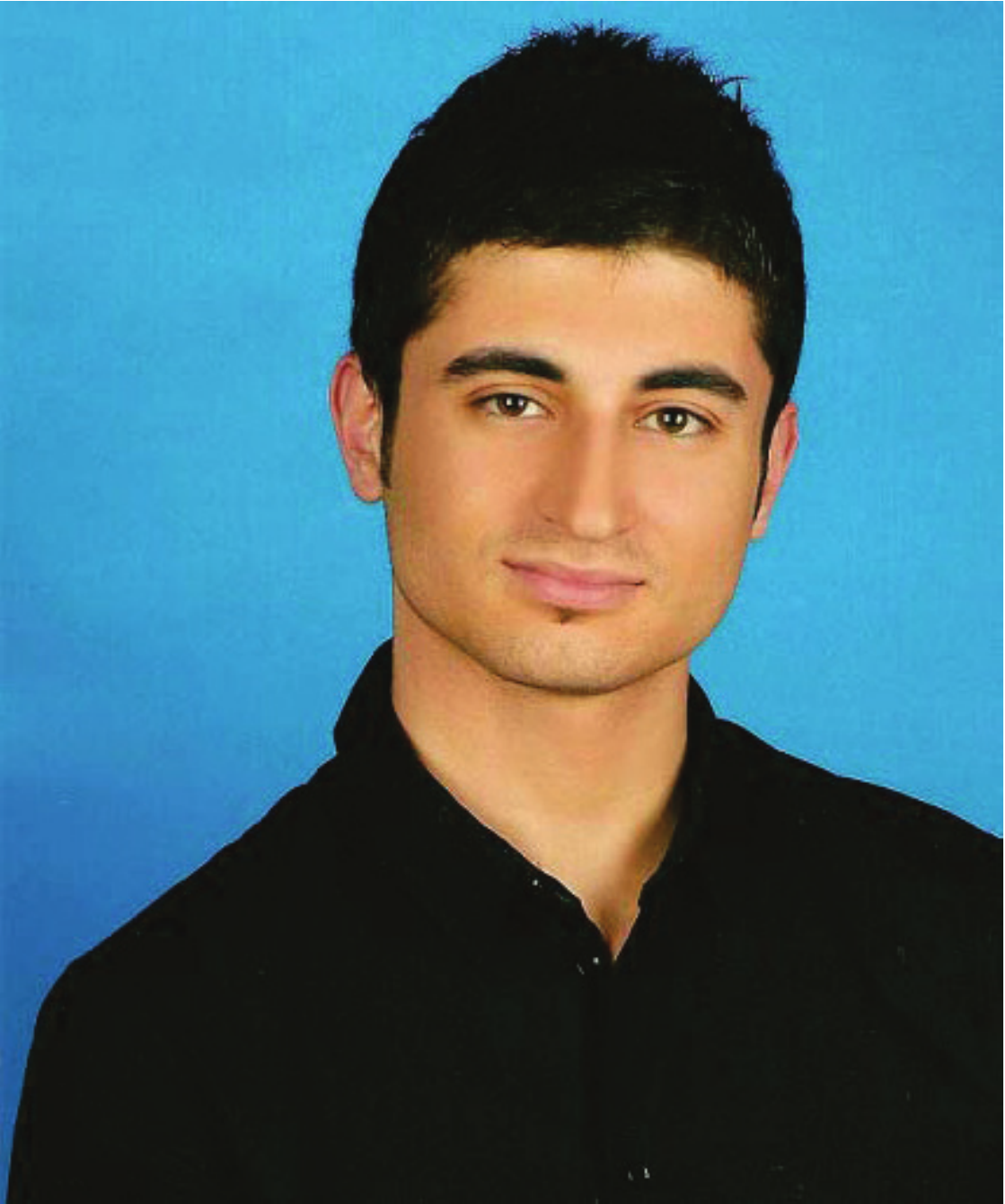}}]{Erkan Kayacan} was born in Istanbul, Turkey, on April 17, 1985. He received the B.Sc. and the M.Sc. degrees in mechanical engineering from Istanbul Technical University, Istanbul, Turkey in 2008 and 2010, respectively. He received the Ph.D. degree in Mechatronics, Biostatistics and Sensors from University of Leuven (KU Leuven), Leuven, Belgium in 2014.

He hold a visiting PhD scholar position in Boston University in 2014 and a Postdoctoral Researcher position in the Delft Center for Systems and Control, Delft University of Technology, Delft, The Netherlands in 2015. He is currently a Postdoctoral Researcher with Coordinated Science Lab, University of Illinois at Urbana-Champaign, USA. His current research interests include control systems, autonomous systems/vehicles and field robotics.
\end{IEEEbiography}

\clearpage
\end{document}